\documentclass[times]{aastex63}
\usepackage[utf8]{inputenc}
\usepackage{amsmath}
\usepackage{enumitem} 
\usepackage{xcolor} 
\usepackage{multirow}
\usepackage[section]{placeins} 
\usepackage{graphicx}
\usepackage{gensymb}
\usepackage{textcomp}

 \received{December 15, 2021 }
 \accepted{March 15, 2022}
\submitjournal{ApJ}
\shorttitle{Infrared Spectroscopic Survey}
\shortauthors{Ali et al.}
\graphicspath{{./}{figures/}}

\begin{document}

\title{A spectroscopic survey of infrared 1-4$\boldsymbol\micro$m spectra in regions of prominent solar coronal emission lines of \ion{Fe}{13}, \ion{Si}{10}, and \ion{Si}{9}}
\correspondingauthor{Aatiya Ali}
\email{aali87@student.gsu.edu}

\author[0000-0003-3196-3822]{Aatiya Ali}
\affiliation{National Solar Observatory,
3665 Discovery Drive, Boulder, CO 80303, USA}
\affiliation{Georgia State University,
25 Park Place, NE \#605, Atlanta, GA 30303, USA }

\author[0000-0002-3491-1983]{Alin Razvan Paraschiv}
\affiliation{National Solar Observatory,
3665 Discovery Drive, Boulder, CO 80303, USA}
\affiliation{High Altitude Observatory,
National Center for Atmospheric Research,
Boulder CO 80307-3000, USA }

\author[0000-0001-8016-0001]{Kevin Reardon}
\affiliation{National Solar Observatory,
3665 Discovery Drive, Boulder, CO 80303, USA}

\author[0000-0001-5174-0568]{Philip Judge}
\affiliation{High Altitude Observatory,
National Center for Atmospheric Research,
Boulder CO 80307-3000, USA }

\begin{abstract}
The infrared solar spectrum contains a wealth of physical data about the Sun and is being explored using modern detectors and technology with new ground-based solar telescopes. One such instrument will be the ground-based Cryogenic Near-IR Spectro-Polarimeter of the  Daniel K. Inouye Solar Telescope that will be capable of sensitive imaging of the faint infrared solar coronal spectra with full Stokes I, Q, U, and V polarization states.
Highly ionized magnetic dipole emission lines have been observed in galaxies and the solar corona. Quantifying the accuracy of spectral inversion procedures requires a precise spectroscopic calibration of observations. A careful interpretation of the spectra around prominent magnetic dipole lines is essential for deriving physical parameters, and particularly, for quantifying the off-limb solar coronal observations from DKIST.
In this work, we aim to provide an analysis of the spectral regions around the infrared coronal emission lines of \ion{Fe}{13} 1074.68 nm, \ion{Fe}{13} 1079.79 nm, \ion{Si}{10} 1430.10 nm, and \ion{Si}{9} 3934.34 nm, aligning with the goal of identifying solar photospheric and telluric lines that will help facilitate production of reliable inversions and data products from four sets of solar coronal observations. The outputs will be integrated in the processing pipeline to produce Level-2 science-ready data that will be made available to DKIST observers.
\end{abstract}


\section{Introduction}\label{sec:intro}

\begin{figure*}[!t]
  \includegraphics[width=0.98\linewidth]{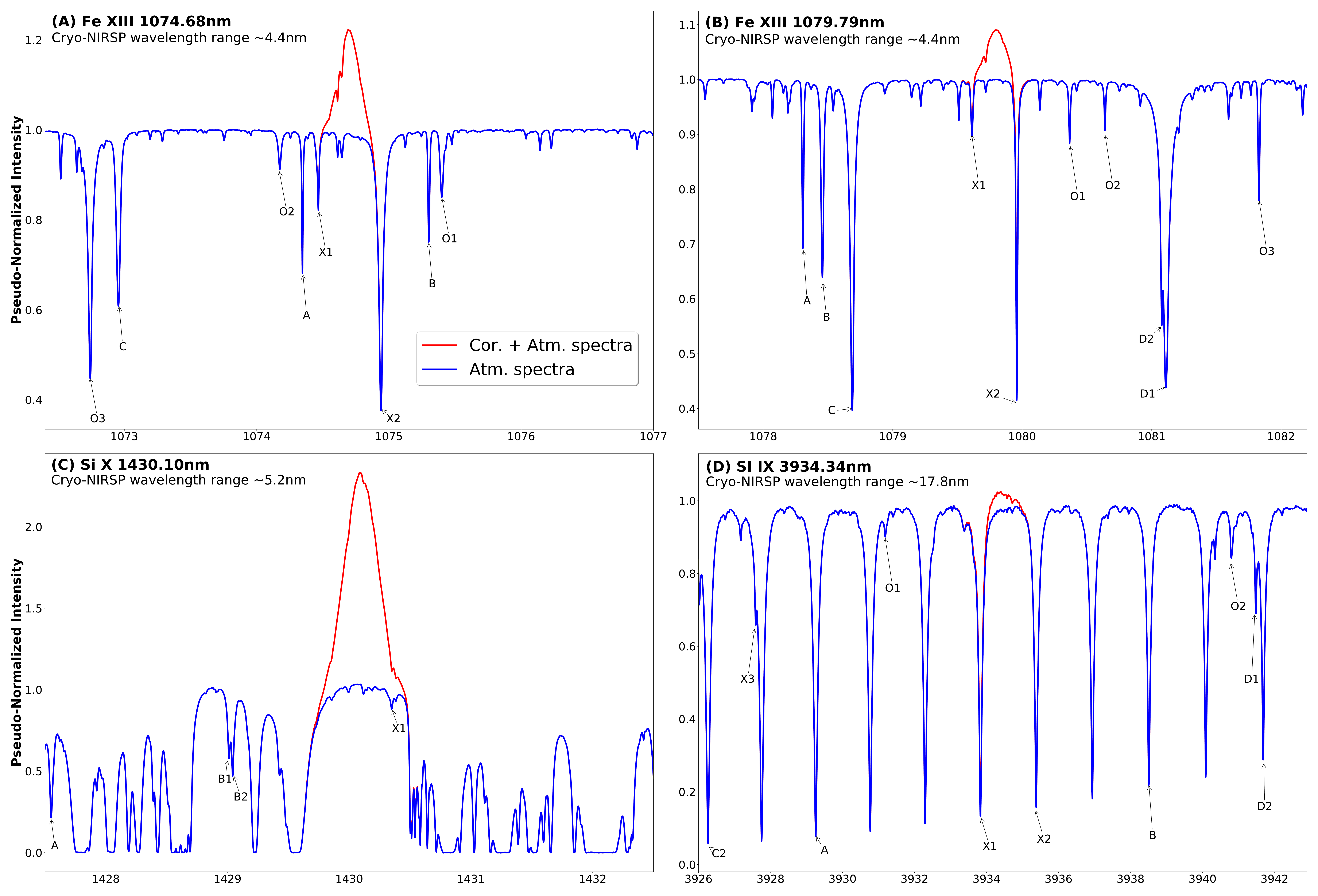}
   \caption{Coronal Line Emission (CLE) synthesized \ion{Fe}{13} , \ion{Si}{10}, and \ion{Si}{9}  emission profiles are plotted over normalized atmospheric absorption spectra. Each coronal ion is centered over a usable {wavelength range }corresponding to a relevant Cryo-NIRSP filter. The theoretical coronal line intensities are {scaled }corresponding to estimations presented in Table \ref{table_ints} {and then added over an atlas of atmospheric absorption spectrum.} The absorption line labels are described in Table \ref{tabtable1}.}
  \label{fig:label3}
 \end{figure*}

Two of the first-light instruments of the Daniel K {Inouye} Solar Telescope \citep[DKIST,][]{2020SoPh..295..172R} will perform spectro-polarimetry of coronal lines in the infrared (IR) region: the Diffraction Limited Near IR Spectro-polarimeter \citep[DL-NIRSP,][]{1987} and the Cryogenic Near-IR Spectro-Polarimeter \citep[Cryo-NIRSP,][]{2017SPD....4811702F}. This work focuses on surveying {select }spectral intervals around prominent coronal lines of \ion{Fe}{13}, \ion{Si}{10}, and \ion{Si}{9}, in the infrared  1-4 $\mu m$ range with the goal of identifying spectral features that can either influence interpretation of coronal measurements from the ground, or be used as references for science calibration. {Figure \ref{fig:label3} shows the four spectral windows of interest relevant to Cryo-NIRSP filters observing the above-mentioned coronal emission lines. We use the Coronal Line Emission (CLE) code \citep[see][for method description]{1999ApJ...522..524C} to synthesize theoretical coronal emission of these lines.}
We aim to identify the effects of scattering and variable telluric absorption features to help obtain accurate physical measurements through the disentanglement of the coronal emission. The results of this analysis are required for developing science-ready ``Level-2'' coronal {observations and} data products.
 

Energetic emitting regions in astronomical sources have traditionally been studied via X-ray, ultraviolet (UV), optical, and infrared emission lines of highly ionized intermediate-mass elements. One of the first correct interpretations of forbidden lines was provided by \citet{1943ZA.....22...30E}. Such lines are often referred to as “coronal lines” since the ions, when produced by collisional ionization, reach maximum abundance at electron temperatures of $\sim$ $10^{5}$ -$10^{6}$ K (or 8.6 eV - 86 eV as measured by \citet{Feldman:77}), typical of the Sun’s upper atmosphere \citep{1993ApJS...88...23G}. 

The emergence of sensitive infrared array detectors prompted existing atomic data to be reexamined for coronal lines sensitive at {expected} field strengths of orders around 10 G  {\citep{kuhnandpenn1995}}. Polarized light from magnetic dipole (M1) lines contain information on coronal magnetic fields in two areas: the direction of the magnetic field projected onto the plane of the sky encoded in the observed linear polarization \citep{1965AnAp...28..877C}, and the strength and sign of the magnetic fields along the line of sight (LOS) encoded in the circular polarization \citep{1969PhDT.........3H,2001STIN...0227999J,2004ApJ...613L.177L}. \citet{1999ApJ...522..524C} present a compact formulation for the description of polarized radiation from M1 transitions occurring in the magnetized solar corona, accounting for the influence of the scattered radiation on atomic polarization induced through both anisotropic irradiation and the Zeeman effect. \citet{2014ApJ...792...23P}, \citet{2020ApJ...889..109D},  \citet{2021ApJ...912...18J}, and \citet{par2021} incrementally build upon each other to {develop} practical methods for inverting magnetic field information from polarized coronal observations. 

The most promising lines lie between 1 and 10 $\micro$m, with the lower limit set by the need to detect small field strengths; and the upper limit by small Einstein A-coefficients and lower intensities of photospheric light \citep{1998ApJ...500.1009J}, with particular emphasis given to astrophysically abundant ions formed in the solar corona (those with {partly filled shells principal quantum numbers} n = 2, 3) with M1 transitions spanning a broad range of wavelengths \citep{2021ApJ...923..186S}. The most promising lines include \ion{Fe}{13} at 1074.68 nm and 1079.79 nm, as well as \ion{Si}{10} at 1430.10 nm, as recorded during the 1998 eclipse and discussed in \citet{1998ApJ...500.1009J}. \citet{1999ApJ...521..478K} further studied the emission line of \ion{Si}{9} at 3934.34 nm, and found consistent results. We chose these IR coronal emission M1 lines as targets for our work.

These lines have been under analysis and speculation for over two decades now for their atomic properties. For example, \ion{Si}{9} has forbidden transitions that hold promise to be diagnostics of coronal magnetic fields \citep{2000ApJ...540.1114B}. Of the M1 lines further studied by \cite{2006ApJ...651.1229J}, those of \ion{Fe}{13} and \ion{Si}{10} are particularly promising because of their relatively strong polarization. All four spectral lines will be observed using Cryo-NIRSP\footnote{\href{https://nso.edu/downloads/cryonirsp_inst_summary.pdf}{Cryo-NIRSP instrument summary document}} where they are part of the main spectral diagnostics to be targeted{, as introduced by \citet{3071}.}

The feasibility of coronal observations of \ion{Si}{10} and \ion{Si}{9} was first established theoretically by \citet{1966ApJ...145..237M}. The radiated power in the lines was predicted by the authors, and found to be large enough to make their detection plausible. Relative to the sky, the brightest predicted emission line observed in the nucleus of a Seyfert galaxy is from the \ion{Si}{9} emission line at 3934.34 nm \citep{2001A&A...369L...5O}. Its detection in the solar corona was one of the primary goals the {airborne IR filter-based imaging observations of \citet{1999ApJ...521..478K}, although the authors were unable to independently confirm the central wavelength of the 3934.34 nm filters to better than a few nanometers.} \citet{1994SSRv...70..185P} hypothesized that the strong \ion{Si}{9} emission line may be the brightest IR coronal emission line to ever be detected.

\citet{2002ApJ...576L.157J} reported a rest wavelength for the line and suggested its potential use as a diagnostic of coronal magnetic fields using the ground-based McMath-Pierce telescope based at the Kitt Peak Observatory \citep{1964ApOpt...3.1337P}. These observations gave the $\lambda_{rest}$ = 3.93434 $\pm$ 0.00007 $\micro m$ where the blue wing of the \ion{Si}{9} emission line overlaps a strong a telluric $N_{2}O$ line{ (see Figure 3 of \citet{2002ApJ...576L.157J}). This observation is matched by us when interpreting the }\ion{Si}{9} calibration candidate lines analyzed in Table \ref{tabtable1}, and portrayed in Figures \ref{fig:label3} and \ref{fig:label9}. {By ``candidates'' we mean absorption lines which we identify herein that can be used to  calibrate velocity shifts and, if needed, to retrieve the un-attenuated coronal spectrum.} Further, these observations show the \ion{Si}{9} 3934.34 nm line being present in active regions of the sun, with intensity levels higher than expected. This may be due to variations in the silicon abundance in the corona, allowing it to remain a strong choice for coronal magnetometry \citep{2002ApJ...576L.157J}. 

Exhibiting linear polarization of magnitude and direction agreeing with resonance polarization theory, measurements produced with the \ion{Fe}{13} 1074.68 nm line by \citet{1967ApJ...150..289E} indicated that the degrees of polarization were somewhat below what was anticipated in a corona with strict radial magnetic fields, {a result confirmed by \citet{arnaud1987}}. Deviations of the field by $20 \degree$ to $30 \degree$ from the radius vector were consistent with measured degrees of polarization. We now know that the degrees of polarization signals can be alternatively explained by fields that might not be in the plane of sky, but cross the integrated volume of a voxel \citep{par2021}. The estimated abundances relative to hydrogen produced in the \citet{1967ApJ...150..289E} analysis (see their fig. 5) agree with iron abundances in the corona.

{Later, \citet{1994ApJ...434..807P} found comparable brightness in the \ion{Fe}{13} lines.} This is also consistent with \citet{1994SSRv...70..185P}, who found \ion{Si}{10} and \ion{Fe}{13} emissions to again be of comparable brightness, although the derived \ion{Si}{10} intensity is higher than ground-based measurements. 

Successive detections of \ion{Si}{10} and of \ion{Fe}{13} were performed by \citet{1996ApJ...456L..67K}, via an airborne experiment during the 1994 eclipse, and ground-based observations obtained by the non-coronagraphic McMath-Pierce solar telescope were also used in this experiment. Sensitive measurements of the near-IR coronal spectrum obtained during this eclipse confirmed again through statistical processes that the \ion{Si}{10} emission line at 1430.10 nm has a brightness comparable to the 1074.68 nm \ion{Fe}{13} emission line. It was hypothesized to be sensitive to magnetic fields 40\% smaller when compared with \ion{Fe}{13} and was consequently proposed as a good choice for coronal magnetometry \citep{1996ApJ...456L..67K}. 

More recently, \citet{2019SoPh..294..166J} surveyed spectra at thermal IR wavelengths and presented broadband polarized light data captured during the 2017 total solar eclipse while testing new technologies for measuring polarized coronal light. 
Spectra of the limb photosphere, chromosphere, prominences, and coronal lines from 310 nm to 5.5 $\mu m$ were all analyzed for this experiment. These observations were taken by an airborne IR coronal imaging spectrometer flown above Kentucky, and a new IR Fourier Transform Spectrometer (FTS) was also deployed from Wyoming. The new FTS offered a particularly high resolution spectra, allowing for the assessment of detailed telluric absorption profiles for new emission lines measured from the aircraft. The aforementioned \ion{Si}{9} line was one of the lines again under analysis. The authors found that the architecture and modernity of the instruments were still inadequate to detect coronal lines from low-altitude sites due to various sensitivities such as atmospheric absorption and background. The Airborne Infrared Spectrometer (AIR-Spec) was constructed with the goal of studying several magnetically sensitive coronal lines, assessing their validity to pursue as future spectro-polarimetric observations \citep{2018ApJ...856L..29S}. The combination of the moderate-resolution FTS and AIR-Spec spectra therefore clearly revealed for the first time the effects of telluric extinction on the IR coronal emission lines. 

The coronal IR spectrum over the range of wavelengths observable by DKIST remains sparsely sampled, and the eclipse of 2017 offered the opportunity to explore this wavelength range in advance of DKIST operations. Motivations secondary to assessing detailed telluric absorptions include searching IR regions for magnetically sensitive coronal lines. Ultimately, AIR-Spec sampled very little of the IR spectra. It is possible that unexpected lines exist in the infrared corona that have not yet been detected. This is why exploring the information in these spectral regions, is a worthwhile goal that will help in the further understanding of the sun's corona.

Emission-line measurements may reveal the presence and nature of current systems in the corona \citep{2006ApJ...651.1229J}, motivating the need to develop instruments capable of measuring polarized light in forbidden coronal lines, such as the DKIST's Cryo-NIRSP. Whether additional ions, e.g. \ion{Mg}{8} $\sim$3028nm, are detectable in ground observations is still debatable. Theoretical intensity calculations and older observational runs are reserved \citep{1966ApJ...145..237M,1971SoPh...21..360O,1998ApJ...500.1009J}, but a definitive ruling has not been reached, as discussed by \citet{2001STIN...0227999J} and \citet{2018ApJ...856L..29S}. For this reason, we limited our coronal line selection to encompass only the more reliable \ion{Fe}{13}, \ion{Si}{10}, and \ion{Si}{9} ions. 

This work is therefore critical to prepare for meaningful observations of coronal M1 lines {on a routine basis}. The necessity mentioned by \citet{1998ApJ...500.1009J} for improved measurements to assess the completeness and accuracy of theoretical spectroscopic work is important to note, aligning with conclusions of needing to perform high resolution spectral analysis {of absorption spectra in proximity of coronal emission}, shown by \citet{2019SoPh..294..166J}.  This kind of analysis can be performed on data like the {airborne} AIR-Spec measurements analyzed by \citet{2018ApJ...856L..29S}, or data obtained from ground-based instruments like the FTS provided via the high-resolution McMath-Pierce atlas, which we utilize here. The aforementioned experiments performed, as well as the data products and properties derived from them, all reaffirm the practicality to inspect the neighboring regions of our four chosen coronal lines. 

The organization of this work is as follows: the architecture and limitations of the instruments and observations used are described in Section \ref{sec:instr}. The analysis procedure and obtained outputs are presented in Section \ref{sec:methods}, followed by the results along with potential issues that can arise in Section \ref{sec:analysis}. Finally, our conclusions are presented in Section \ref{sec:summ}. The detailed spectroscopic analysis and best fits for each of the different candidates of \ion{Fe}{13}, \ion{Si}{10}, and \ion{Si}{9} absorption lines are included in Appendix \ref{sec:app}.

 \section{Observations{, Instrumentation, \& Modeling}} \label{sec:instr}

{DKIST is an off-axis capable telescope. Its Cryo-NIRSP instrument is designed as coronagraph. It was designed to take full advantage of this capability by primarily observing the 1-5 $\mu m$ range, a region of low scattered light with enhanced sensitivity to the Zeeman effect.} The telescope configuration provides the ability to conduct unobstructed off-axis observations. Cryo-NIRSP is mounted on the DKIST Coud\'{e} platform and will measure the full polarization state (Stokes I, Q, U, and V) solar spectral lines at wavelengths ranging between $\sim$ 1000 nm - 5000 nm, using one of two slits: one of length 233\arcsec and width 0\farcs5 with a spectral resolving power of $\sim$30000; and the second slit of length 120\arcsec and width of 0\farcs15 with a spectral resolving power of $\sim$100000. Effective pixel platescales and resolutions will be dependent on the combination of slit and filter in use \citep[see][]{2010SPIE.7735E..8CN,3071}. {For this work, we considered the most updated instrumental parameters in \href{https://nso.edu/downloads/cryonirsp_inst_summary.pdf}{[1]} and quote resolutions of 0.027 nm, 0.027 nm, 0.033 nm, and 0.107 nm (lower spectral res. 0\farcs5 slit) along with 0.008 nm, 0.008 nm, 0.011 nm, and 0.036 nm (higher spectral res. 0\farcs15 slit)  corresponding to the \ion{Fe}{13}, \ion{Si}{10}, and \ion{Si}{9} wavelength ranges, respectively.}

To harness the multitude of advantages of operating at IR wavelengths, Cryo-NIRSP must be cryogenically cooled to $\sim$ 70 K to reduce any background noise that could compromise the faint solar coronal signal \citep{2017SPD....4811702F}. Cryo-NIRSP is the DKIST instrument with the dedicated capability of sensitively imaging the relatively faint IR corona and thermal IR solar spectrum. Its polarimetric and spectrographic capabilities will allow measurements of the faint coronal emission with unprecedented cadence, resolution, and accuracy. 

In the current facility configuration, Cryo-NIRSP can not work in conjunction with other DKIST instruments and does not utilize the DKIST Wavefront Correction (WFC) system. Thus, increased uncertainties arising due to mixing of different spatial positions may manifest, especially in the case of longer integration times {and the narrower $0\farcs15$ slit.}

We also briefly mention the DKIST DL-NIRSP diffraction grating based integral field spectrograph, that has a low resolution mode of 0\farcs93 in the 500 nm – 1800 nm wavelength range \citep{1987}. Our analysis is also relevant for DL-NIRSP as our selected lines of \ion{Fe}{13} and \ion{Si}{10} might be simultaneously observable by two of the three bands (900 nm – 1350 nm and 1350 – 1800 nm) with full Stokes polarization. 

{Our four selected coronal line emission profiles are synthesized using CLE and scaled based on theoretical or observed literature intensities. The available literature has debated over which conditions coronal line intensities can be measured from ground instrumentation, airborne instrumentation or via balloon flights \citep[e.g.][]{1994ApJ...434..807P,1999ApJ...521..478K, 2004ApJ...613L.177L, 2014LRSP...11....2P,2018ApJ...856L..29S,2019SoPh..294..166J,2021arXiv210509419S}. Theoretical calculations for coronal line intensities at heights of $\sim$1.1$ R_{\odot}$ were computed by \citet{1998ApJ...500.1009J}, and adapted by us in Table \ref{table_ints}. Theoretical intensities for infrared lines including those of coronal ions have been reviewed more recently by \citet{2018ApJ...852...52D} where the authors stress that the atomic data requires improvements. We note for these strongest lines the conclusions of \citet{1998ApJ...500.1009J} currently remain valid.}

\begin{deluxetable}{ccccc}
\tablehead{\colhead{Cor. line} & \colhead{Theoretical} & \colhead{Observed} & Exp.& \colhead{Source} \\[-0.2cm]  
\colhead{}&\colhead{Int.}&\colhead{ Rel. Int.}&\colhead{Time}&\colhead{}}
\tablecaption{A compilation of relative {to background} coronal line intensities derived from literature observations. Ground, eclipse, flight and balloon campaigns are considered. Due to the limited number of intensity calibrated observations available in the literature, we show only relative to background values. Theoretical{ intensity }calculations at $\sim$1.1 $R_{\odot}$ are adapted from \citet[Fig. 3]{1998ApJ...500.1009J} and are in units of $erg~cm^{-2}~s^{-1}~st^{-1}$. }
\label{table_ints}
\startdata
          &       & 1.2 &40& Fig. \ref{fig:japanspec}\\
Fe XIII   & 22.4 & 1.4 &5& \citet{2009ScChG..52.1794B}\\
1074.68nm &       & 1.5 &40-70& \citet{2002PASJ...54..807S}\\
          &       & 4.6 &120& \citet{2014LRSP...11....2P}\\
&&&&\\
          &       & 1.2 &40& Fig. \ref{fig:japanspec}\\
Fe XIII   & 5.2  & 1.1 &5& \citet{2009ScChG..52.1794B}\\
1079.79nm &       & 1.1 &40-70& \citet{2002PASJ...54..807S}\\
          &       & 2.4 &120& \citet{2014LRSP...11....2P}\\
&&&&\\
Si X      &5.4   & 2.3  & 5 &\citet{2018ApJ...852...23D}\\
1430.10nm &       & 4.4 &120&\citet{2014LRSP...11....2P}\\
&&&&\\
Si IX     &0.7   &1.0  &300&\citet{2014LRSP...11....2P}\\
3934.34nm &       &      &   &
\enddata
\end{deluxetable} 

{At this time, Cryo-NIRSP is still being integrated into the DKIST facility, and there is no available data that can be used for the purposes of this work. Our spectral line identifications will be performed using an ``atlas'' of high spectral resolution calibrated observations compiled from the FTS instrument of the McMath-Pierce telescope \citep{1996ApJS..106..165W,ftsdpw}. The FTS is able to obtain ground-based spectral data by reconstructing the interference pattern of vacuum frequency observations. The FTS is capable of obtaining spectra from approximately 22 $\mu m$ to 0.7 $\mu m$, with resolving power and resolution limited by the total path difference obtainable with the spectrometer. The FTS atlas spectra covers all of our four regions of interest with resolutions of 0.0004 nm, 0.0004 nm, 0.0013 nm, and 0.0080 nm respectively for each coronal ion wavelength range}.

In order to validate and test our FTS atlas identifications, we utilized observed coronal spectra encompassing the two \ion{Fe}{13} lines in the 1074 nm - 1080 nm IR regions, as analyzed by \citep{2002PASJ...54..807S} using observations done at the Norikura Observatory. The coronal spectral observations were performed with a {25cm aperture refracting coronagraph}. A CCD camera mounted on the system, has a pixel size of 24 $\mu$m x 24 $\mu$m with a 1024 x 1024 format allowing for the observation of a 6 nm portion of the spectrum. {The achieved spectral resolution was $\sim 0.018$ nm, which is comparable with the Cryo-NIRSP values listed above.} The pixel resolution of the spectroheliograms after binning the CCD readout was approximately 2$\arcsec$ x 2$\arcsec$ but the slit width that was equivalent to 4$\arcsec$ on the Sun, which restricted the spatial sampling to about 2$\arcsec$ x 4$\arcsec$ \citep{2002PASJ...54..807S}. 

 \section{Methodology} \label{sec:methods} 
 
 Understanding the origins and magnitude of the contamination of ground-based observations would overall aid in the interpretation of observed coronal data and prepare it for data analysis and assist in the automation of the processing of the data acquired by DKIST. {When discussing polarised observations, we focus here on interpreting the Stokes I component, as Stokes Q, U, and V are formally just derivatives of I \citep{1999ApJ...522..524C}.}  The goal for DKIST coronal Level-2 data processing is to invert physical coronal information from raw Level-1 Stokes profiles obtained using Cryo-NIRSP coronal line observations. Addressing neighboring {photospheric} absorption and telluric {absorption} lines will further reduce uncertainties introduced through the observations, and provide guidance on how to reduce {and reliably calibrate }observations of these same lines in more broad galactic applications and possibly in future stellar coronae observations.

\subsection{Searching for Photospheric \& Atmospheric Lines Suitable for Interpreting Coronal Emission} \label{sec:ghost}

We start by searching for{ photospheric and telluric }absorption profiles in proximity to the 1074.68 nm, 1079.79 nm, 1430.30 nm, and 3934.34 nm wavelength ranges using the FTS Kitt Peak spectral atlas. The FTS atlas was corrected for wavelength conversions from the vacuum to air values in all three spectral windows of interest using the approximation provided by \citet{1996ApOpt..35.1566C}. By proximity, we mean we selected windows, of the size of the spectral range covered by the Cryo-NIRSP {prefilters}, in which each of the four \ion{Fe}{13}, \ion{Si}{10}, and \ion{Si}{9} lines, were centered in the respective interval. The widths of the spectral intervals are 4.4 nm for the \ion{Fe}{13} 1074.68nm and 1079.79nm lines, 5.2 nm for \ion{Si}{10} 1430.10 nm, and 17.8 nm for \ion{Si}{9} 3934.34 nm.
 
We then assessed which absorption lines to treat as candidates by utilizing the BAse de donn\'{e}es Solaire Sol\footnote{\href{http://bass2000.obspm.fr/solar\_spect.php}{BASS2000 survey spectra}} (BASS2000) high resolution solar spectrum database \citep{BASS2000}. Each potential absorption line was searched for in the database to ensure that it was not a spurious effect portrayed as a phantom dip in the spectra. In addition, we did not consider "weak" lines, defined herein as having a drop of less than 0.1 in relative intensity (column 7 in Table \ref{tabtable1}) in the FTS atlas. The lines {can} saturate {in their cores}, making identification of their centroid difficult, were also not considered. The absorption features were then verified using the BASS2000 database and were subsequently deemed usable in the case of positive identification, or deemed unusable by not finding a corresponding profile in BASS2000. This procedure also served as a double check for the confidence of the FTS atlas.

The National Institute of Standards and Technology (NIST) Atomic Spectra Database\footnote{\href{https://physics.nist.gov/PhysRefData/ASD/lines\_form.html}{NIST atomic spectral data query form}} \citep[ASD;][]{NIST_ASD} was then used to assess the air wavelengths for the chosen {photospheric} absorption lines from the spectra, and to subsequently identify them. The identifications usually matched the centroid wavelengths of the FTS atlas. In some cases, minimal differences on the order of $<$ 0.01 nm between the ASD and the FTS atlas were found. These differences might become significant in the context of achievable {wavelength calibrations}.
Further, the ASD contains information on the relative line intensity and the transition probability $A_{ki}$ that was utilized to confirm that the respective transition is indeed the one responsible for our observed absorption profile. 

{So far, the focus has been on photospheric absorption profiles. We now shift our attention to potential molecular telluric candidate absorptions.} Ground-based observations using the non-coronagraphic McMath-Pierce telescope \citep{2002ApJ...576L.157J} have provided observed spectra of {telluric molecular absorption lines} in the IR regions. Therefore, when identifying molecular absorptions that would not be otherwise covered by the NIST ASD database, we utilized \citet[Figure 1]{2014LRSP...11....2P} \citep[originally presented in ][]{2003AAS...203.3803H} that catalogs atmospheric transmission in the IR 0.5 - 5.5 $\mu$m range, to single out plausible molecules. We then utilized the high-resolution transmission (HITRAN) molecular absorption database\footnote{\href{https://hitran.org/}{HITRAN molecular spectra query form}} \citep{HITRAN2017} to confirm our identification as follows: The observed air wavelengths, confirmed using BASS2000, that could not be associated with any {NIST ASD atomic} lines were converted to vacuum wavenumbers, and then queried with the HITRAN output of the singled-out molecules in each desired wavelength range. All HITRAN findings are obtained with main isotopologue molecular sequences, with an abundance $>0.99$. The HITRAN line intensity S and Einstein Coefficient A (analogous to ASD's $A_{ki}$) are utilized to confirm the identification of each suitable line. The HITRAN wavenumbers are then converted back to air wavelengths. Similar differences on the order of $<$0.01 nm between HITRAN data and the FTS atlas are reported in a limited number of cases. HITRAN wavelength calculations and intensities are presented in Table \ref{tabtable1} for identified lines of H$_2$O \citep{H2O-nu-44-762,H2O-nu-60-1208,H2O-nu-30-15,H2O-nu-44-762}, N$_2$O \citep{N2O-nu-6-206}, CH$_4$ \citep{CH4-S-36-744,CH4-nu-33-239}, and SO$_2$ \citep{SO2-nu-13-1352} molecules.

Shown in Figure \ref{fig:label3} are the \ion{Fe}{13}, \ion{Si}{10}, and \ion{Si}{9} selected spectral ranges in which various candidate absorption lines were identified and analyzed using statistical processes discussed in Section \ref{sec:analysis}. {The pseudo-normalized units here reflect the FTS atlas being normalized {to a local continuum value of unity}, and the coronal synthesized spectra intensities {being} scaled {relative to this continuum} and added to the background counts. The relative to background coronal intensities shown are calculated using estimations collated in Table \ref{table_ints}, that are then added to background spectra.} Currently, we {have very few measurements of the relative line} emission with respect to the background. This is further {complicated} by the variation in the background intensity, which depends on the amount of scattered light from the sky and telescope optics. An example of an actual \ion{Si}{9} observation {that is the "weakest" line in our set,} is presented in \citet[Fig, 3]{2002ApJ...576L.157J}. {Our estimation of the \ion{Si}{9} intensity in panels D of Figures \ref{fig:label3} and \ref{fig:shifts} also provide an example of how intensity might manifest in practical observations.} 

The line identifications are recorded in {Table \ref{tabtable1}. The identification procedures are described in Section \ref{sec:candidates}. The properties of all the absorption candidate lines are shown for each spectral window corresponding to one coronal line, as labeled in the first column. The second column records the identifying labels for each analyzed absorption profile. The labelling of the lines identified in Table \ref{tabtable1} follow three categories: alphabetical (e.g. A, B, C, D1, D2, etc.), ``O'' lines (e.g. O1, O2, etc.), and ``X'' lines (e.g. X1, X2, etc.), with respect to each of the four selected wavelength ranges. The {alphabetical} category consists of lines identified using the FTS atlas and confirmed with BASS2000 and the NIST ASD database or identified using HITRAN. The second category are lines considered as back-up candidates that might be used in certain conditions, while the last category of lines were those that were found problematic to interpret. {Such problems may involve:} saturated intensity counts; very weak absorption profiles; overlap with the coronal emission lines centered in the spectral window assuming no Doppler shifts are occurring; {or when lines become} uncertain when convolving the FTS atlas with Cryo-NIRSP instrumental specifications, etc. The implications of Table \ref{tabtable1} are expanded in Section  \ref{sec:linesets}. }

The comments might describe lines that do not have a clear corresponding atomic or molecular emission at those wavelengths, or in some cases, lines that are portrayed by ASD or HITRAN to correlate to a transition that should not be strong enough, or to an atom that does not have a minimal solar abundance (i.e. \ion{Th}{1} accidentally corresponding to 1074.17 nm, etc.). Given these discrepancies in identification of some of these lines, not all are considered by us suitable for disentangling the coronal emission. {Appendix Figures \ref{fig:label6}, \ref{fig:label7}, \ref{fig:label8}, and \ref{fig:label9} show the detailed spectral positions and characteristics of all the absorption line identifications presented in Table \ref{tabtable1}, that are corresponding to the spectral windows associated to each of our four coronal lines of interest.}

 \subsection{Spectral Line Fitting} \label{sec:math}
 
 \begin{figure}[!h]
    \centering
     \includegraphics[width=0.5\columnwidth]{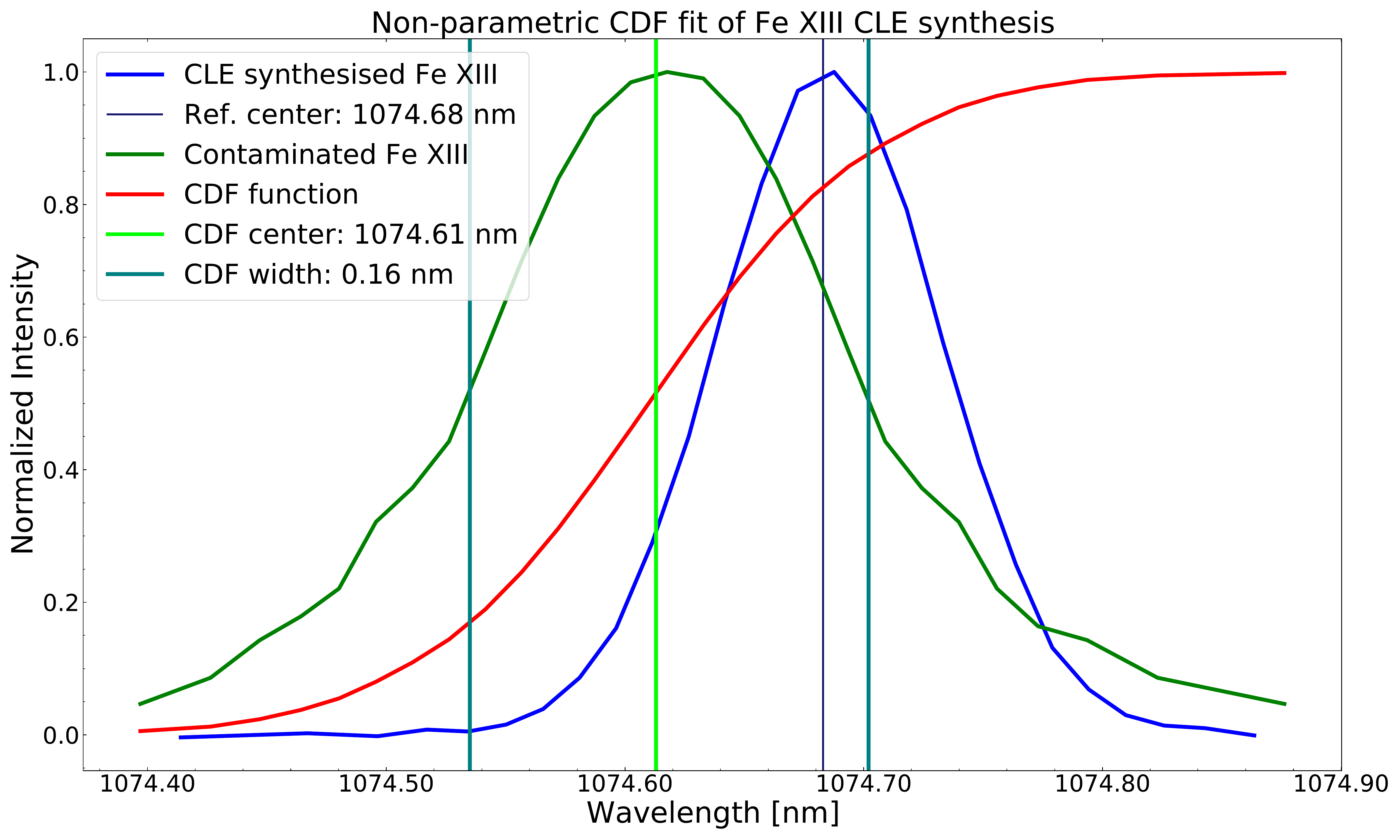}
    \caption{\centering CDF non-parametric fitting applied to synthetic CLE Stokes I coronal spectra for the \ion{Fe}{13} 1074.68 nm line. The blue emission profile represents the synthesized theoretical atomic emission, while the green profile depicts the same profile after it has been contaminated by wavelength shifts and profile broadening due to atmospheric air column movements and turbulence. {The Red curve represents the CDF fit of the contaminated \ion{Fe}{13} line.}}
    \label{fig:cdf}
\end{figure}  

We employ a non-parametric method, using the Cumulative Distribution Function (CDF) to pre-process {the coronal Stokes I profiles and lines identified in the FTS atlas. This simple, yet robust statistical method allows general line properties to be quantified using intensity measurements (see example in Figure \ref{fig:cdf}). The method is more general than standard Gaussian fitting as it enables us to sum data as a function of wavelength, and interpret it in a statistical fashion. The fitting method is used herein on both coronal emission and absorption lines from spectral data. In practice, we take} the main line parameters resulted from the CDF fit and used them as initial parameters for a curve fitting algorithm, in order to obtain sub-pixel accuracy. A meaningful interpretation requires the assumption that the Stokes I profiles are compatible with a normal distribution. {The achievable spectral resolution of the Cryo-NIRSP spectrograph will also cause most profiles to be similar to normal distributions.} Under the limitations of our data and for the goals of this work, we therefore consider this assumption to be satisfied.

Statistical methods applied to the CDFs of Stokes intensity vs. wavelength can be used to derive the properties of the lines, where the wavelengths corresponding to the 16\%, 50\%, and 84\% of the normalized distribution are recovered. {For a true normal distribution, this 16-84\% range corresponds to $\sim2\sigma$, where $\sigma$ correlate to the standard deviation of a normal distribution.}

Parameters that can be recovered using this approach include wavelength shifts, and from it, Doppler velocities, as well as line widths, and uncertainties. In particular, wavelength shifts can then be calculated from the difference between the distribution centroid wavelength (50\%) and a laboratory value. In Figure \ref{fig:cdf}, {Stokes I profiles of \ion{Fe}{13} coronal emission in ideal (pure) and noise influenced conditions (contaminated by turbulence which would mix different spatial locations in the same pixel, lowering the spatial resolution and mixing multiple profiles) are synthesized by us using CLE.} Table \ref{table_lines} presents a comparison between CDF and Gaussian fitting, where both perform almost identically in the idealized case. The non-parametric approach is seen to be more flexible in more complex cases.

\begin{deluxetable}{lcc}[t]
\tablehead{\colhead{} & \colhead{Contaminated \ion{Fe}{13}} & \colhead{Pure \ion{Fe}{13}}}
\tablecaption{Example of non-parametric CDF and parametric Gaussian fitting for CLE a synthesized \ion{Fe}{13} 1074.68 nm observation. The pure synthesized observation is compared to an uncertainty contaminated version of itself. The CDF analysis is applied to the contaminated spectra. The Gaussian width here is 2$\sigma$, and the CDF width is defined as the difference between 84th and 16th percentiles.}
\label{table_lines}
\startdata
CDF 50th (Median) & 1074.61 & 1074.68 \\
CDF 84th & 1074.69 & 1074.74 \\
CDF 16th & 1074.54 & 1074.63 \\
CDF Width [nm]& 0.16 & 0.11  \\
Gauss Mean & 1074.61 & 1074.68 \\
Gauss Width [nm]&0.14 & 0.11 \\[-0.95cm]
\enddata
\end{deluxetable} 

{In general, our method aims to isolate coronal emission lines in observed spectra, extract their main parameters via CDF analysis, then use these to perform a detailed curve fit to finely tune these extracted parameters. We then search for usable absorption profiles in the same wavelength range, perform the same type of fitting on these, and then look for differences when comparing them with a laboratory/database reference positions, enabling us to use such differences to calibrate the coronal line.}
In order to extract the parameters using this CDF based approach for our absorptions of interest identified in Section \ref{sec:ghost}, we selected small spectral windows so that the line under scrutiny was isolated without any other absorption influences. Some lines were very hard to isolate; e.g. the spectral range around the 1430.10 nm \ion{Si}{10} emission as shown in Figure \ref{fig:label8}. Some combinations of two adjacent lines are also selected and documented in Table \ref{tabtable1}. These are fitted with a custom CDF approach suited for multi-peak identification.

\subsection{Validating the Spectral Identification and Fitting} \label{sec:candidates} 

\begin{figure}[!ht]
    \centering
    \includegraphics[width=0.5\textwidth]{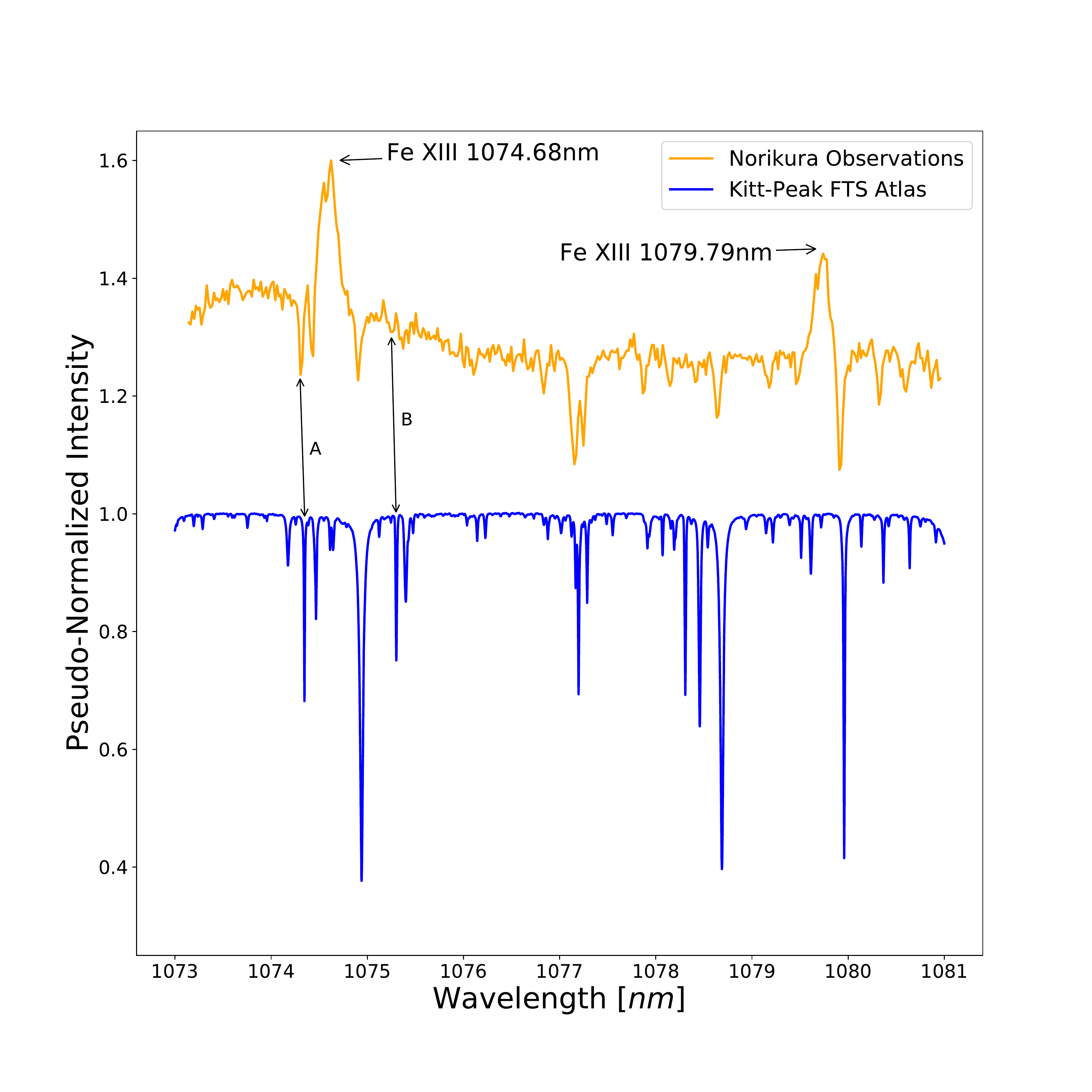}
    \vspace{-0.6cm}\caption{\centering Norikura Observations compared to the Kitt Peak FTS atlas in the \ion{Fe}{13} spectral window against a pseudo-normalized intensity. These spectra were compared in order to determine the best choices for the candidate lines for analysis for the \ion{Fe}{13} coronal line. Here, the Norikura Observations are slightly shifted vertically to clearly visualize both spectra. {The labels A and B identify two absorption lines for the \ion{Fe}{13} 1074.68 nm coronal line, as observed by the Norikura Observatory. The individual line fits are expanded in Figure \ref{fig:fe1lines}.}}
    \label{fig:japanspec}
 \end{figure}

Raw observational spectra from the Norikura Observatory were used in conjunction with the identifications corresponding to the two \ion{Fe}{13} lines to observationally validate the survey identifications, and show that these absorption lines can be used in ground observations to calibrate coronal emission. {The identified photospheric and atmospheric absorptions in Table \ref{tabtable1} are compared to the raw spectral observations.} In practice, these raw Norikura observations were checked against the FTS atlas, where observed central wavelengths and depth of absorptions were then compared to the identified lines.{ A {visual} comparison between raw Norikura spectra and the corresponding FTS atlas subrange is shown in Figure \ref{fig:japanspec}. Figure \ref{fig:fe1lines} presents problems that might appear when trying to utilize these atlas candidates with actual observations. This is expanded in Section \ref{sec:compare}.}

\section{ Results and Discussion }\label{sec:analysis}

\subsection{Coronal Line Intensities}\label{sec:intensity}
{Figure \ref{fig:shifts} depicts examples of expected observation for the  four coronal lines under scrutiny, showing estimated coronal intensities and how typical coronal LOS outflow velocities influence their spectroscopic observations.  {Additionally, we did not normalize the integration time to {counts per second} for the relative intensities in Table \ref{table_ints} due to not being able to distinguish in all sources whether the line cores are saturated due to instrumental or physical effects.} The {theoretical intensities} adapted in Table \ref{table_ints}, although being inconclusive by themselves, raise compelling questions that can be pursued via DKIST Cryo-NIRSP observations. }

The {observed relative intensities, adapted via Table \ref{table_ints}, corroborate with the findings of \citet{2018ApJ...856L..29S} resulting from eclipse observations taken above the atmosphere. Additional weaker coronal ions like \ion{Mg}{8} at 3027.64 nm were detected from those altitudes. We bring to attention Figure \ref{fig:shifts} (D) showing that ``expected" intensities for the \ion{Si}{9} line are weak with respect to the background and close to detection limits, based on the current available observations and theoretical intensity estimations. This leads us to consider \ion{Si}{9} as representing a detection baseline for this selection. We did not further provide estimations for other scientifically interesting coronal lines like the \ion{Mg}{8} line mentioned above that are the target of Cryo-NIRSP due to the uncertainty of high-confidence detection from the ground, even for an instrument like Cryo-NIRSP.} Thus, we consider our four discussed coronal lines to represent the best candidate choices for coronal spectro-polarimetry and magnetic field diagnostics, as they are strong enough to be observed from the ground with next generation instrumentation like Cryo-NIRSP.

{Due to the yet inconclusive nature of current observations, we chose to present the Appendix Figures \ref{fig:label6}, \ref{fig:label7}, \ref{fig:label8}, and \ref{fig:label9} as an overlap of CLE synthesized coronal emission and the FTS atlas spectra, using just pseudo-normalized units where coronal emission is scaled arbitrarily between 1 and 2, without considering the above discussed intensity calculations.}

{We reaffirm the need for improvements needed in atomic and theoretical intensity calculations, as correlating different coronal observations proved difficult due to difference in instrumentation, observation conditions, locations, height matching, integration times, and scarce absolute intensity calibrations.}

\subsection{Comparing Atmospheric Absorption Spectra with Coronal Observations}\label{sec:compare}

\begin{figure*}[b]
    \includegraphics[width=1\linewidth]{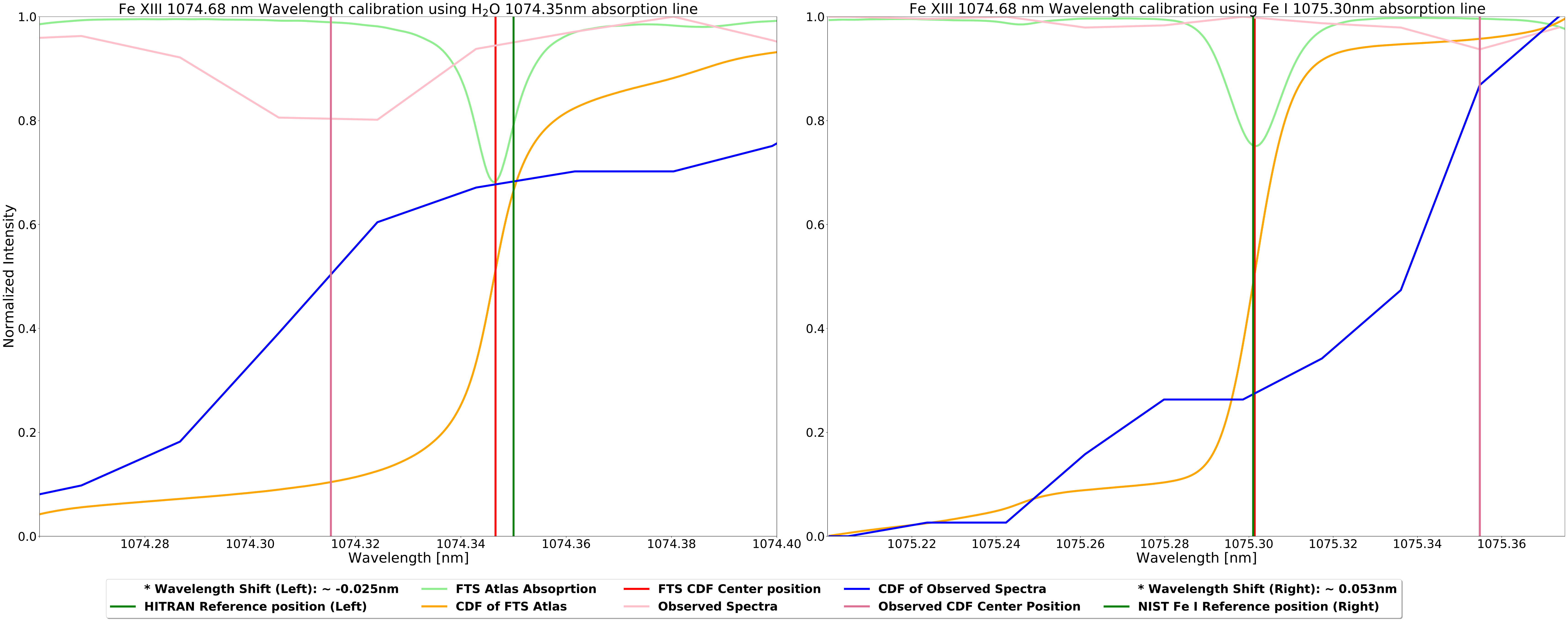}
    \caption{Identified calibration lines are used to calibrate the \ion{Fe}{13} 1074.68 nm coronal line. The FTS atlas spectra is plotted with the raw observation, revealing atmospheric shifts that will influence or even compromise coronal observations. {Left Panel:} \ion{Fe}{13} wavelength calibration using the H$_{2}$O 1074.35 nm absorption line. {Right Panel:} Incorrect wavelength calibration while using the "B" \ion{Fe}{1} 1075.30 nm photospheric absorption line identified in Table \ref{tabtable1}. The line could not be isolated in the particular time instance of the raw observation, leading to a bad fit. The effect can be corrected by coupling multiple identified lines in the analysis.}
    \label{fig:fe1lines}
 \end{figure*}

Figure \ref{fig:japanspec} plots both the FTS atlas spectra and Norikura spectra analyzed by \citet{2002PASJ...54..807S} for one spectral position at $\sim$1.1 R$_{\odot}$, in the range covering both \ion{Fe}{13} coronal lines. In Figure \ref{fig:fe1lines} (left), we portray the results of the analysis applied to one example absorption line in the \ion{Fe}{13} spectral window. The selected line, centered at $\sim$ 1074.35 nm could not be identified in the ASD database.{ Instead, a} HITRAN H$_2$O candidate line was confirmed. The FTS atlas is plotted with the raw observation over the same spectral range. Both of these had the CDF analysis applied to them. The center positions of the distributions are marked and were calculated from the distribution statistics as described in Section \ref{sec:math}. The spectral sampling of the raw observation is significantly coarser than that of the FTS atlas. Taking this into account, we show that the CDF method does recover the centroid of the absorption, with an uncertainty mainly given by the spectral sampling of 0.0187 nm. We draw attention to the spectral positions between the two spectra, highlighting a 0.041 nm shift that {is present in }the raw observation. This shift likely indicates an error in the definition of the absolute wavelength scale of the Norikura data, which depends on orbital motions, solar rotation, and instrumental alignment and drifts. This is important in context as modern instruments will achieve resolutions of orders of magnitude more significant; e.g. Cryo-NIRSP will reach pixel resolutions in the order of 0.009 nm. In addition, we take note that in this case, a small difference exists between the FTS atlas CDF line center and the theoretical ASD or HITRAN air wavelength. This small 0.004 nm difference, {correspondent to a $\sim$1 $\mathrm{km\,s^{-1}}$ Doppler velocity,} might be due to the uncertainty of the spectral sampling of the FTS atlas, or by a not-as-accurate identification of the ASD air wavelength. Either way, this difference is negligible towards our effort, as it is {smaller than what even a next generation instrument like Cryo-NIRSP can resolve as measuring such a difference in the highly broadened coronal lines will be challenging.}

All the candidate calibration lines for both \ion{Fe}{13} coronal lines were scrutinized in the same way, revealing and confirming the same atmospheric shifts, as for the purpose of this example we only used one time instance in the raw observation. We could not obtain similar raw observations in the \ion{Si}{10} and \ion{Si}{9} spectral windows, as these are significantly less studied.

In Figure \ref{fig:fe1lines} (right) we show a more convoluted example of an identification, also in the \ion{Fe}{13} spectral window. As recorded in Table \ref{tabtable1}, the absorption line B at 1075.30 nm is a candidate selected for analysis, identified as a photospheric \ion{Fe}{1} {absorption} line in the ASD. We applied the CDF methods to both the FTS atlas and observed spectra as in the above example, and found that in this instance the observed spectra's CDF peak was at $\sim$ 0.055nm to the right of the ASD and FTS atlas central wavelengths. This immediately indicated that there was a problem during the semi-automated analysis of the data, since this is the same instance of the observation as in Figure \ref{fig:fe1lines} (left), subjected to the same atmospheric conditions. Although higher resolution spectra from the FTS atlas and BASS2000 allowed for the identification of the line with a reasonable 0.75 absorption dip, the line was too weak in the raw observation to permit recovery. Thus, the procedure incorrectly fitted an unrelated line in the sampled spectral range. In order to avoid such problems, we set up our procedure to fit multiple absorption lines in conjunction, (e.g. A, B, and C) across the \ion{Fe}{13} spectral window, comparing the independent determinations, and automatically rejecting outliers.

\begin{deluxetable*}{cccccccl}
\movetableright=0.7cm
\rotate
\tablehead{ \colhead{Cor. Line \& Wave.} &\colhead{ID}&\colhead{Cal. Line}&\colhead{Source}&\colhead{Wave.}&\colhead{CDF Wave. Range}&\colhead{Rel. I}&\colhead{Comment}}
\tablecaption{Candidate absorption lines for atmospheric shift calibration as respective to the four spectral windows corresponding to the chosen coronal lines. Not all candidate lines could be identified in the NIST ASD database. The Wavelength column follows the FTS atlas, HITRAN or ASD air wavelength when they match. In the case of small differences between the two or when a HITRAN or ASD identification could not be made, the FTS atlas wavelength is quoted. Some absorption lines could not be separated, and should be analyzed as a pair. Numbered letters eg. O1,O2,... or X1, X2,... are less desirable due to weak or saturated absorption profiles, overlap the main coronal lines, etc.}
\label{tabtable1}
\startdata
\multirow{7}{44pt}{~~~Fe XIII 1074.68nm} & O3 &  Si I  & ASD & 1072.74nm & 1072.58nm - 1072.92nm & 0.44 & Very strong line; too close to slit wavelength margin\\
        & C  &  C I & ASD  & 1072.95nm & 1072.84nm - 1073.12nm & 0.60 & Strong absorption line\\
        & O2 & H$_2$O & HITRAN & 1074.15nm & 1073.96nm - 1074.32nm & 0.91 & Very weak line; {Cryo-NIRSP detection uncertain}\\
        & A & H$_2$O & HITRAN & 1074.35nm & 1074.24nm - 1074.44nm & 0.68 &  Strong absorption. Blend of two water lines\\
        & X1 & Fe I & ASD & 1074.46nm & 1074.40nm - 1074.52nm & 0.82 &  Weak line, unusable due to blend with cor. \ion{Fe}{13} line\\   
        & X2 &  {Si I}  & ASD & 1074.94nm & 1074.80nm - 1075.10nm & 0.38 & Strong line, unusable due to blend with cor. \ion{Fe}{13} line\\
        & B & Fe I  & ASD & 1075.30nm & 1075.20nm -  1075.37nm & 0.75 & Not very prominent but might be usable\\
        & O1 &  C I  & ASD & 1075.40nm & 1075.31nm - 1075.46nm & 0.85 & Weak line; possible blend issue \\
\hline
\multirow{11}{44pt}{~~~Fe XIII 1079.79nm}& A & Fe I & ASD & 1078.30nm & 1078.20nm - 1078.40nm & 0.69 & Strong absorption line\\
        & B  &  Si I & ASD & 1078.45nm & 1078.32nm - 1078.58nm & 0.63 & Strong absorption line\\
        & C  &  {Si I}& ASD & 1078.68nm & 1078.48nm - 1078.96nm & 0.39 & Very strong line\\
        & X1 & H$_2$O & HITRAN & 1079.60nm & 1079.56nm - 1079.70nm & 0.89 &  Weak line. Strong blend with cor. \ion{Fe}{13} line\\      
        & -  &  - & - &     -     & - & - & Unusable due to blend with cor. Fe XIII line\\
        & X2 & H$_2$O & HITRAN & {1079.95}nm & 1080.32nm - 1080.40nm & 0.41 & Strong line. Blend of two water lines\\
        & -  &  - & - &     -     & - &  -  & Unusable due to blend with cor. Fe XIII line\\
        & O1 & H$_2$O & HITRAN & 1080.37nm & 1080.16nm - 1080.56nm & 0.88 & Weak line; Weak blend with  cor. \ion{Fe}{13} line; {Cryo-NIRSP detection uncertain} \\
        & O2 &  Ne I & ASD  & 1080.63nm & 1080.56nm - 1080.72nm & 0.91 & Very weak line; {Cryo-NIRSP detection uncertain}\\
        & D2 & H$_2$O & HITRAN & 1081.08nm & 1080.72nm - 1081.52nm & 0.56 & Strong absorption 2-line set. {Weak for Cryo-NIRSP lower specral res.}\\
        & D1 & Mg I & ASD  & 1081.11nm & 1080.72nm - 1081.52nm & 0.43 & Strong absorption 2-line set; treat as pair\\   
        & O3 & Fe I & ASD  & 1081.83nm &1081.64nm - 1082.04nm & 0.77 & Moderately strong line; close to slit wavelength margin\\
\hline
\multirow{4}{44pt}{~~~~Si X 1430.10nm} & A & Fe I & ASD & 1427.50nm & 1427.48nm - 1427.62nm & 0.21 & Separable line in tough range; needs specialized fitting procedure\\
        &B1 & H$_2$O & HITRAN & 1429.01nm & 1428.83nm - 1429.14nm & 0.47 & Pair of close lines, fit together; {Weak for Cryo-NIRSP lower spectral res.}\\
        &B2 & H$_2$O & HITRAN & 1429.04nm & 1428.83nm - 1429.14nm & 0.57 & Blend with B1\\
        &X1 & Fe I & ASD & 1430.30nm & 1430.29nm - 1430.34nm  & - & Not usable due to  blend with cor. \ion{Si}{10} line; {Cryo-NIRSP detection uncertain} \\
\hline
\multirow{11}{44pt}{~~~~Si IX 3934.34nm} &  {C1} & CH$_4$ & HITRAN & 3926.02nm & 3925.48nm - 3927.08nm & 0.71 & {Weak for Cryo-NIRSP lower spectral res.} couple with N$_2$O band \\      
        &{C2} &N$_2$O & HITRAN & 3926.26nm & 3925.48nm - 3927.08nm & 0.06 & Molecular absorption to couple with C1 \\
        &  {X3} & CH$_4$ & HITRAN & 3928.68nm & 3928.62nm - 3928.72nm & 0.67 & Not usable due to proximity to N$_2$O molecular band \\
        & A & N$_2$O & HITRAN & 3929.25nm & 3928.12nm - 3930.28nm & 0.08 & Molecular absorption band \\
        & {O1} & N$_2$O & HITRAN & 3931.17nm & 3931.00nm - 3931.40nm & 0.90 & Weaker N$_2$O line; {Cryo-NIRSP detection uncertain}\\
        & X1 & N$_2$O & HITRAN & 3933.83nm & 3933.32nm - 3936.52nm & 0.13 & Absorption band; not usable due to  blend with cor. \ion{Si}{9} line \\
        & X2 & N$_2$O & HITRAN  & 3935.38nm & 3933.32nm - 3936.52nm & 0.15 & Absorption band; not usable due to  blend with cor. \ion{Si}{9} line\\
        & {B} & N$_2$O & HITRAN & 3938.50nm & 3938.12nm - 3938.92nm & 0.22 & Molecular absorption band\\  
        & {O2} & SO$_2$ & HITRAN & 3940.80nm & 3940.44nm - 3941.24nm & 0.84 &  Weak line. Weak S in HITRAN. Blend with weak water line\\          
        & {D1} & CH$_4$ & HITRAN & 3941.47nm & 3941.16nm - 3942.12nm & 0.69 & Blend of two methane lines. Couple with N$_2$O band\\  
        & {D2} & {N$_2$O} & HITRAN & 3941.66nm & 3941.16nm - 3942.12nm & 0.29 & {N$_2$O} Molecular band for D1\\
\enddata
\end{deluxetable*}

\subsection{Sets of Potential Calibration Lines for Coronal Observations} \label{sec:linesets}

 Every absorption candidate line's spectral position and label, {as recorded in Table \ref{tabtable1},} can be visualized in Figures \ref{fig:label6}, \ref{fig:label7}, \ref{fig:label8}, and \ref{fig:label9}. {The high resolution spectra for each selected candidate line can be found in the four subsections of appendix \ref{sec:app}, corresponding to each of the four spectral windows in Figures \ref{fig:appfe1}, \ref{fig:appfe2}, \ref{fig:appsi1}, and \ref{fig:appsi2}. The CDF method is applied to the absorption spectra for each scrutinized line, and their centers are compared to their NIST ASD or HITRAN reference positions.} Note that the absorption profiles of these panels are reversed into emission for the ease of reading each plot.

The alphabetical category of labeling {in the ID column} encompasses lines that have deep absorption profiles, with a {minimum residual intensity of $<$0.8.} For most of them, we could identify the corresponding atom in the ASD or the HITRAN atmospheric molecular absorptions by using \citet{2014LRSP...11....2P}. In general cases, these lines can be isolated in the spectra, but more convoluted situations exist. In some cases, multiple lines affected by blending need to be analyzed as a pair; e.g. D1 and D2 in the spectral window of \ion{Fe}{13} 1079.79 nm. Additionally, all lines in this category are not close to the edge of our chosen spectral windows.

The second category of lines {encompasses those where one} or more of the criteria described for the above category is not satisfied. For example, we use Figure \ref{fig:label6} and scrutinize line O3, finding it to have very strong absorption, but is too close to the {left end of the Cryo-NIRSP prefilter wavelength range} to routinely use in the analysis process. In the same spectral window, lines O1 and O2 are both weak lines. The ``O'' lines can be weaker than 0.8 residual intensity, but we have not chosen candidates weaker than 0.91. As shown in Figure \ref{fig:fe1lines} (right), weaker lines might not be reliable by themselves in non-ideal conditions.

The third category of lines {were those that were either very weak or overlapping the {not-Doppler shifted} coronal emission lines within the spectral range.} These lines were identified, labeled, and quantified in order to understand and mitigate the effects of the underlying spectra on coronal emission, but were deemed weaker candidates due to uncertainties caused by this blending. In Figure \ref{fig:label6}, lines X1 and X2 overlap and blend with the \ion{Fe}{13} coronal emission. We note that our synthesized \ion{Fe}{13} emission encompasses only thermal effects that dictate the broadening. In an actual observation, the emission will probably be significantly broader, principally due to unresolved plasma motions along the LOS \citep{1966gtsc.book.....B}. These lines become usable in situations where significant Doppler shifts are observed, as shown in Figure \ref{fig:shifts}.  

{Figures \ref{fig:label6}, \ref{fig:label7}, \ref{fig:label8}, and \ref{fig:label9} also present the effects of applying a convolution of the atmospheric spectra with a Gaussian function having a full width at half maximum (FWHM) of the Cryo-NIRSP {nominal spectral resolution in} the wavelength range for each selected coronal line. This showed that all the labels defined using the criteria above remain valid {even for actual} Cryo-NIRSP observations.}

Column four then documents the sources that were utilized to identify each absorption line. The two sources are the ASD for atomic lines or HITRAN for molecular lines. The air wavelengths, as recovered from one of the sources, were cross-checked with the FTS atlas and BASS2000. The resulting wavelengths are then recorded in column five.

The calculated relative intensities are determined using the FTS atlas and represent the difference of the absorption peak from the normalized background of the spectra. Brief, specific notes per candidate line made are also recorded on the 7$^{th}$ column of Table \ref{tabtable1}. 

\subsection{Absorption Wavelength Range and Accuracy}

Column six of Table \ref{tabtable1} records usable ranges for CDF fitting for all discussed absorption profiles. Some closely spaced lines (as described in the comments) might benefit from a multi-peak fitting approach, which requires computing derivatives of the CDF profiles. Alternatively, the profiles can also be directly fitted with more standard parametric function approaches inside the same wavelength ranges and Double-Gaussian functions can be used where required.
Based on the ASD and HITRAN identification and/or \citet{2014LRSP...11....2P} analyses, we separated the identified line into photospheric absorption and terrestrial molecular absorptions. The interpretation for atmospheric based molecular absorptions like N$_2$O in the \ion{Si}{9} spectral window is straightforward. Atmospheric lines from molecules of N$_2$O, H$_2$O, CH$_4$, and SO$_2$ are considered to be very stable. The accuracy in using such lines for calibrations has been established by \citet{1985A&A...149..357C} to be in order of 0.05 $\mathrm{km\,s^{-1}}$. The base estimation was more recently confirmed by \citet{2010A&A...515A.106F}, where additional corrections can improve the accuracy to up to 0.02 $\mathrm{km\,s^{-1}}$. In our context, deviations from the rest wavelength are interpreted as being due to shifts in telluric columns and the profile broadening or narrowing caused by a change in local opacity. Solar photospheric lines like \ion{Fe}{1}, \ion{Si}{1}, \ion{C}{1}, etc. require a different interpretation. These should exhibit roughly weak Doppler velocities, corresponding to photospheric movements on the order of 0.5-1 $\mathrm{km\,s^{-1}}$ \citep[e.g.][]{2003ASPC..293..197A,2011ApJ...740...15R,2013ApJ...765...98W} where transitory flows of the order of $\sim5$ $\mathrm{km\,s^{-1}}$ might occur  \citep{2008ApJ...677L.145N}. These movements will {generally }cancel-out as in such cases, the light {in this background is derived from the atmospheric} scattering of full-disk contributions of photospheric light. Orbital motions of the Earth might also influence these lines. 

On the other hand, coronal outflows can have velocities in the order of tens to hundreds of $\mathrm{km\,s^{-1}}$. For example, recent spectroscopy of coronal jets revealed outflow speeds in between 17-170  $\mathrm{km\,s^{-1}}$\citep{2021A&A...647A.113Z}. In the cases of large impulsive ejecta, like Coronal Mass Ejection (CMEs) outflows can routinely reach velocities of over 1000 $\mathrm{km\,s^{-1}}$ in extreme examples \citep[see][ and references therein]{2004ApJ...604..420Z,2011ApJ...738..191B}. Additionally, in open coronal regions, propagation of acoustic waves is $\sim$150 $\mathrm{km\,s^{-1}}$, and typical Alfv\'{e}nic wave propagation velocities are on the order of 400-600 $\mathrm{km\,s^{-1}}$ \citep[see][]{2015NatCo...6.7813M}. Not all observational cases will encompass outflowing structures. There is significant interest in analyzing more subtle motions in the corona on the order of 1-5 $\mathrm{km\,s^{-1}}$ \citep[e.g.][]{2007Sci...317.1192T}. For this exercise all velocities should be considered upper limits as we did not take into account projection effects.

Figure \ref{fig:shifts} shows how typical {LOS outflow speeds} can influence a spectral observation. For this, we overlapped the FTS atlas spectra with a CLE simulation of coronal emission with encoded {LOS outflow speeds} of $\pm$5, $\pm$20, $\pm$50, $\pm$150, and $\pm$400  $\mathrm{km\,s^{-1}}$. For three of the four panels, two versions of the plots are provided.{ The highest LOS speed is only plotted for panel (D) as it is the only region where the Cryo-NIRSP slit is wide enough to recover such speeds.}

The top plots show the Doppler shifted coronal lines with the pseudo-normal intensity units described in Section \ref{sec:intensity}. We observe how certain absorption lines might become detectable against the shifted or broader/narrower coronal emission line profiles. In other words, lines that are normally not recommended might be used as good calibration candidates in certain conditions{, thus the inclusion in the list.} Taking this effect into account, we advise that any calibration line selection method should dynamically select the most optimal calibration lines, based on the raw spectral position of the coronal line center. 

The bottom plots show calculated intensities that simulate the observed relative intensity values provided in Table {\ref{table_ints}}. The weakest intensities were selected as corresponding to each coronal line, as to provide a cautious overview of how strongly these lines manifest in a real observation. As can be seen especially in panel D, a very cautious and meticulous pre-processing is needed in order to recover the coronal emission with respect to the background, {which is expected to vary considerably depending on local conditions at the observatory.  }

The wavelength ranges of the long Cryo-NIRSP slits for the \ion{Fe}{13} and \ion{Si}{10} ions will be insufficient to analyze very fast moving coronal LOS velocities or Alfv\'{e}nic waves as described above, when centering the slits on the rest wavelengths of the ions in question. As dynamically changing diffraction grating {angle} with the Cryo-NIRSP is not currently provisioned, \ion{Si}{9} is a recommended choice {for such studies} due to the significantly larger slit spectral {coverage}. We can observe in panel D, that in this case, interpretation is hindered by the low predicted and observed coronal signal, but also due to the very deep and closely spaced N$_2$O band absorption lines.     

We finally remind the reader that when dealing with single vantage point observations, e.g. Cryo-NIRSP observations, any observed velocity might be only a small fraction of a true velocity due to projection effects. If we thus assume a required velocity accuracy in coronal measurements to be on the order of $\sim 1$ $\mathrm{km\,s^{-1}}$, this will lead to a wavelength calibration accuracy of $\sim$ 0.004 nm at 1074.68 nm, and $\sim$ 0.013 nm at 3934.34 nm. The future commencement of DKIST observations will put this requirement to the test.

\section{ Summary and Future Prospects \normalfont}\label{sec:summ}

This work focused on surveying {certain narrow windows in} the infrared spectra around 1-4 $\mu$m to identify suitable spectroscopic lines that {detrimentally and systematically change the coronal data, but which } can be used to calibrate highly ionized coronal magnetic dipole lines emitted by solar but also by galaxies. {The extreme dimness of stellar coronae compared with photospheres  make such lines undetectable in the absence of high quality coronagraphs occulting regions very close to the photosphere, something not yet achieved in exo-planetary imaging.} The resolution and quality of the observed spectra used in this paper is in some cases below the quality of data that will be attained by modern large aperture instruments like Cryo-NIRSP, nevertheless our analysis and results of absorption spectra offer a baseline list of wavelengths calibrations that can help to interpret coronal emission observed by facilities such as the novel DKIST.

To summarize:

\begin{figure*}[p]
    \centering
    \includegraphics[width=1\linewidth]{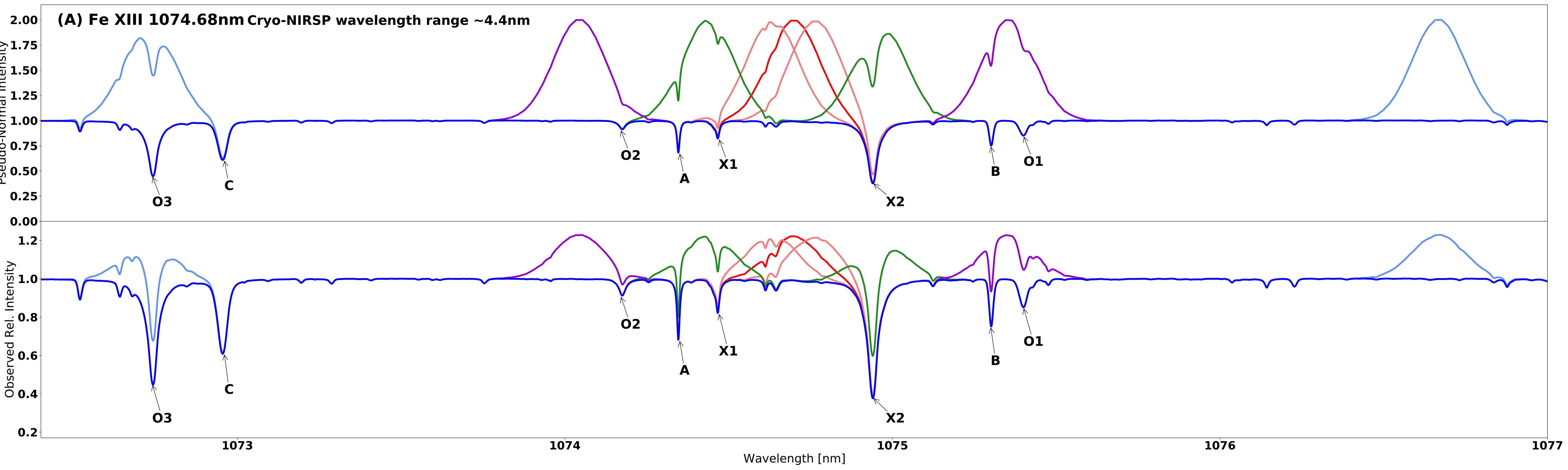}
    \includegraphics[width=1\linewidth]{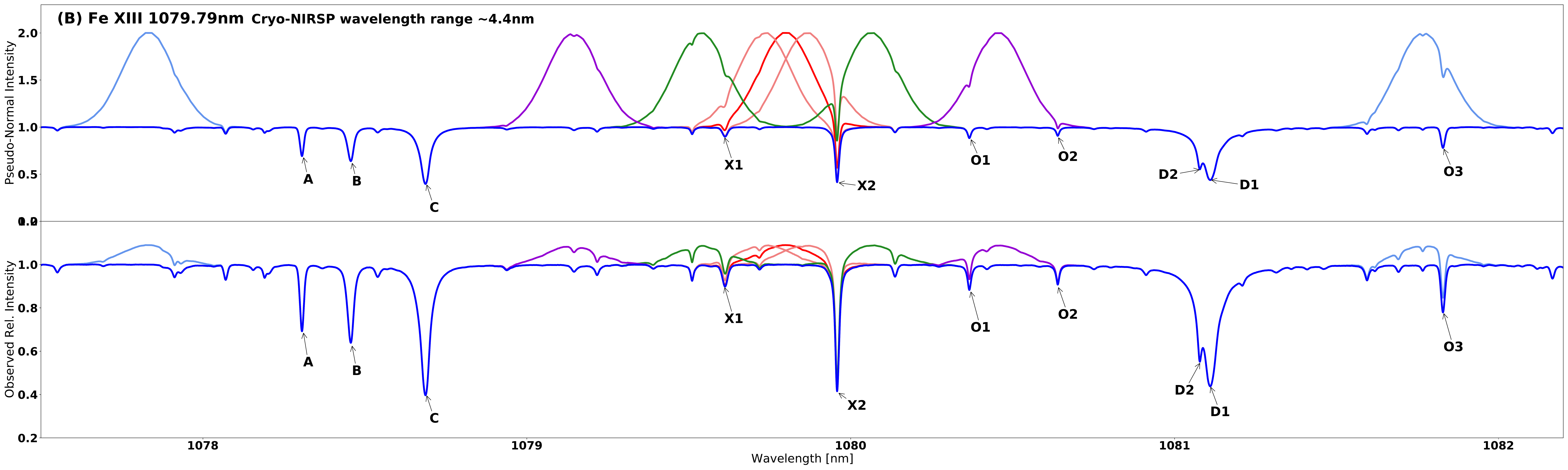}    
    \includegraphics[width=1\linewidth]{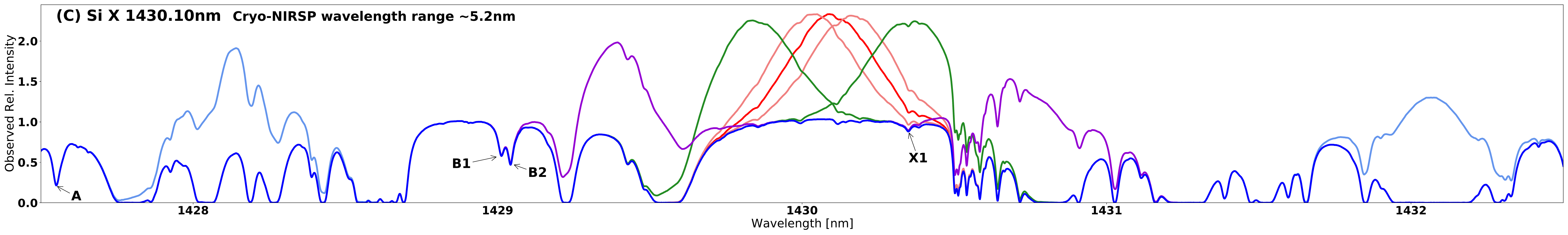}
    \includegraphics[width=1\linewidth]{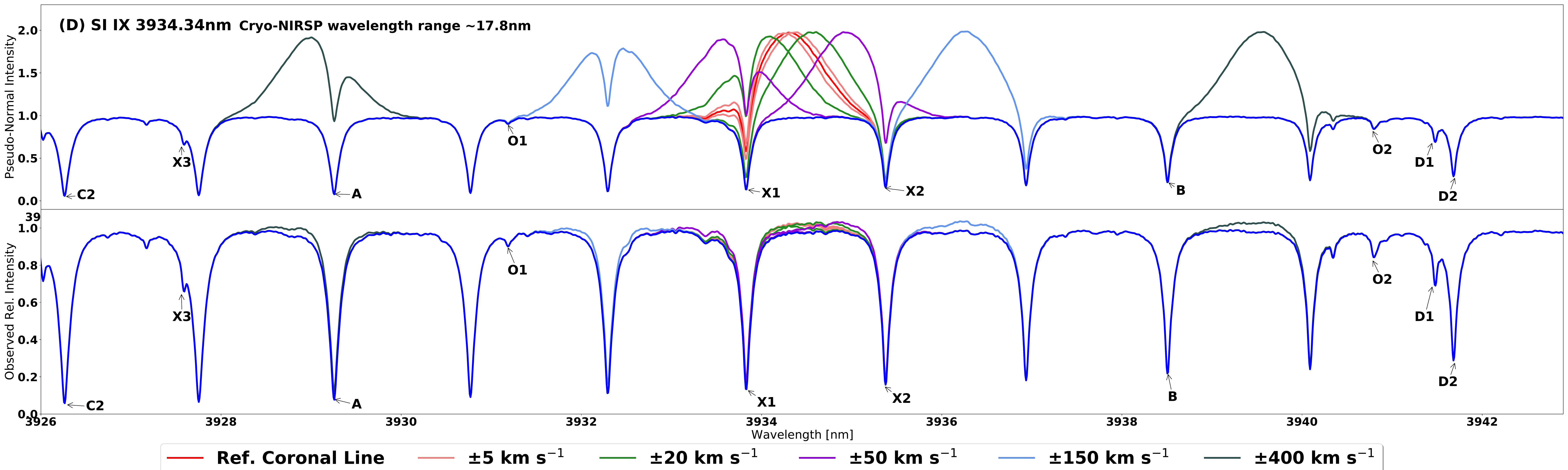}
    \caption{Multiple blue-red Doppler shifts of $\pm 5$, $\pm 20$,$\pm 50$,$\pm 150$, and $\pm 400$ (panel D {only}) $\mathrm{km\,s^{-1}}$ are applied to the simulated coronal emission and then {combined} with the FTS atlas spectra, over wavelength ranges equivalent to the relevant Cryo-NIRSP {prefilters}. For panels A, B, and D, plots using both pseudo-normal counts (top) and observationally derived relative intensities (bottom) and are provided. The weakest intensities from Table \ref{table_ints} are chosen. In the case of panel C, only the observed relative intensity is plotted because counts are similar in both interpretations. In the case of panel D, The wavelength is sufficiently large to analyze high-speed streaming coronal plasma, although, as can be seen, the background relative intensity will complicate interpretation.}
    \label{fig:shifts}
\end{figure*}  


$\bullet$ Modern instruments like Cryo-NIRSP will observe IR solar coronal data with a sensitivity never obtained before. The instrument can not use adaptive optics corrections, making it prone to contamination effects that need correcting before quantitatively analyzing observations.

$\bullet$ To this purpose, this work focuses on analyzing spectral windows in the IR in proximity to the outstanding solar coronal lines of \ion{Fe}{13} 1074.68 nm, \ion{Fe}{13} 1079.79 nm, \ion{Si}{10} 1430.10 nm, and \ion{Si}{9} 3934.34 nm. 

$\bullet$ We identify and analyze absorption profiles using the NIST ASD or HITRAN databases in the relevant spectral windows corresponding to our four chosen coronal emission lines, and confirm the results by cross-checking the BASS2000 survey spectra.  

$\bullet$ We identify lines of both solar (photospheric) and atmospheric absorption (molecular line), and quantify their properties. Additionally, we discuss that in the context of expected coronal {LOS Doppler shifts}, the identified  lines can be used to provide accurate values of coronal Doppler shifts. 

$\bullet$ {Our categorizing of the candidates from their absorption spectra provides new information that can be utilized to adequately quantify the {spectral properties of }coronal IR emission lines of interest. {We note that {absolute and relative} intensities remain an open question and can not be constrained with the methods described here.}}

$\bullet$ The validation for the absorption profiles in the \ion{Fe}{13} spectral range via the comparison with raw observations provided a basic understanding of the varying observational conditions that could influence the future observations DKIST will perform.

$\bullet$ {Although our CDF based fitting method proved reliable, our testing} of the calibration procedure showed that in poor observing conditions, some profiles might not be accurately recovered, leading to errors in the analysis. {We mitigate this by simultaneously using the combination of all clearly identifiable absorption candidate lines when performing a coronal calibration.}

$\bullet$ The relative intensity calculations derived from previous observations show that some of the coronal lines in question might be hard to disentangle from background signal.

$\bullet$ {Strong absorption profiles and in the case of \ion{Si}{10} low transmission manifest in the spectral windows of interest. The influence of such identified} absorption lines on coronal emission is not negligible. {This is further complicated when} dealing with outflowing plasma. DKIST observations might, in some cases, require advanced planning, including additional calibration data, when expecting highly Doppler shifted spectra. 

$\bullet$ Quantifying the level of contamination that could affect observations and enhance the recovery of spectral properties and results from coronal observations is a goal not covered in this work. Our future aim is to use these spectroscopic results to assess the quality of a potential coronal observation.  

The results presented in Table \ref{tabtable1} are of particular importance when preparing data for addressing and inverting coronal magnetic field information - a goal of DKIST's  Cryo-NIRSP and DL-NIRSP instruments. The same issues will need to be addressed for accurately resolving magnetic fields in future larger field-of-view (FOV) instrumentation, like COSMO \citep{2018IAUS..335..359T} and UCoMP \citep{2019shin.confE.131T}, both of which will provide novel observations that can augment DKIST data.

\acknowledgments
{The authors thank the anonymous reviewer for her/his comments that greatly helped us improve this work.} We additionally thank D.A. L\u{a}c\u{a}tu\c{s} for the careful reading and review of the initial submission. BASS2000 Data have been obtained thanks to the IR solar spectrum obtained from Kitt Peak observations. BASS2000 website has been used to get those data. Data from the NIST atomic spectra database queried via the NIST ASD website. The Norikura observation used by us as a validation dataset was provided via courtesy of T. Ichimoto.

A.A. was funded by the National Science Foundation (NSF) Boulder Solar Alliance REU program, award \#1659878. A.R.P. and K.R. were funded by the National Solar Observatory (NSO), a facility of the NSF, operated by the Association of Universities for Research in Astronomy (AURA), Inc., under Cooperative Support Agreement number AST-1400405. P.J. was funded by The  National  Center  for  Atmospheric  Research, sponsored  by  the  National Science Foundation. 

\bibliographystyle{aasjournal}
\bibliography{bibliography}{}

\appendix
\vspace{-0.8cm}\section{Spectroscopy and CDF fitting for all lines discussed}
This section contains detailed spectroscopic plots of the spectral regions of interest and the best fits for each candidate of the \ion{Fe}{13}, \ion{Si}{10}, and \ion{Si}{9} absorption lines. Each spectral region is also individually convolved with a Gaussian function correspondent to the Cryo-NIRSP 0\farcs5 slit spectral resolution, as dependent on the appropriate filter. In addition, we show the individual CDF fits of all identified photospheric and atmospheric absorptions.
\label{sec:app}
\subsection{\ion{Fe}{13} 1074.68 nm spectral window}
 
  \begin{figure}[h]
   \centering
   \includegraphics[angle=270,origin=c,width=0.62\linewidth]{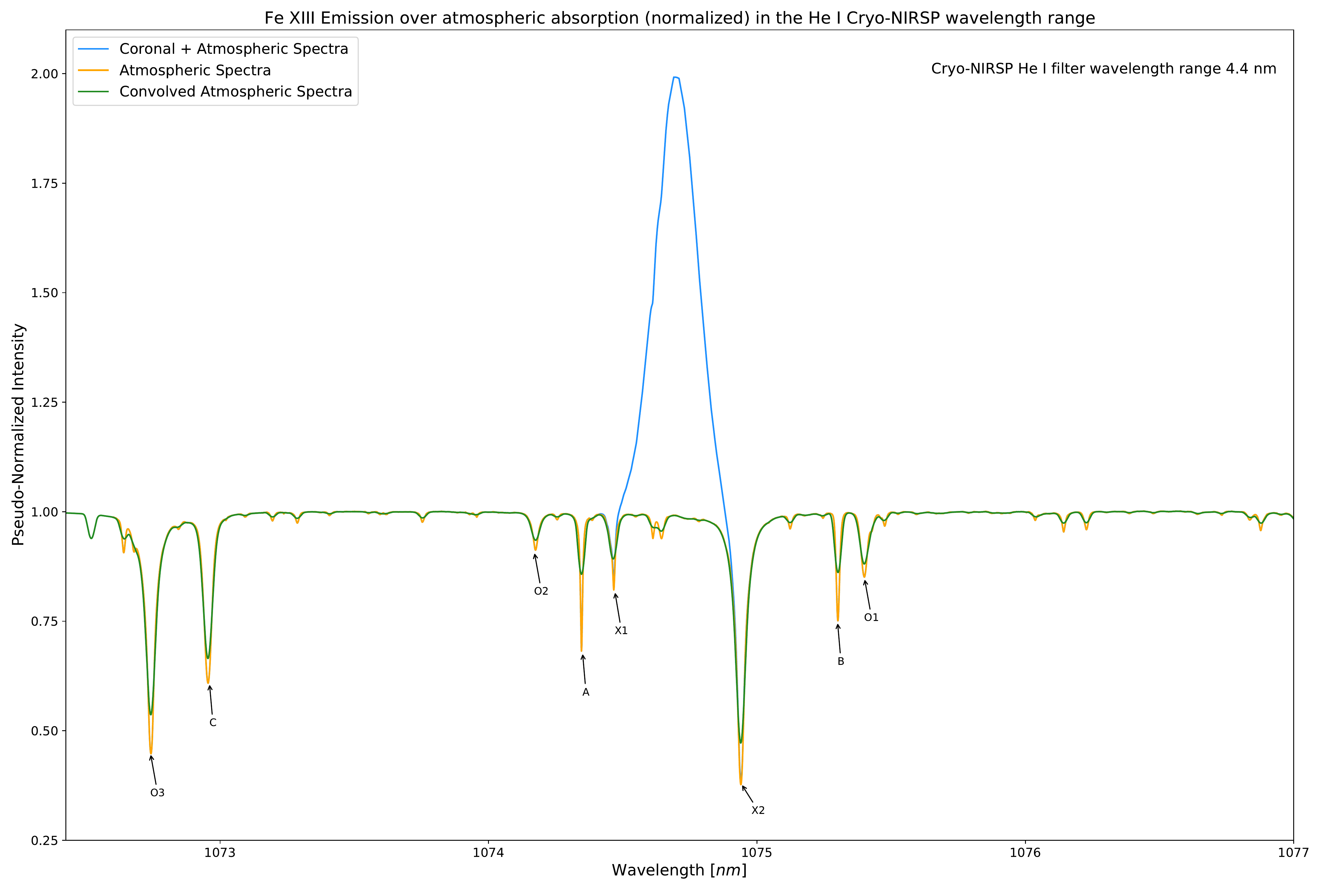}
   \caption{\centering \ion{Fe}{13} 1074.68 nm emission over atmospheric absorption (normalized) in the He I Cryo-NIRSP wavelength range with all labeled candidate lines in Table \ref{tabtable1}. The green curve represents the atmospheric spectra that is convolved with a Gaussian function correspondent to the Cryo-NIRSP 0\farcs5 slit spectral resolution of 0.027 nm of the \ion{He}{1} filter (see \href{https://nso.edu/downloads/cryonirsp_inst_summary.pdf}{[1]}).}
   \label{fig:label6}
 \end{figure} 
 
 \FloatBarrier 
 
\begin{figure}[!hb]
  \includegraphics[width=0.34\linewidth]{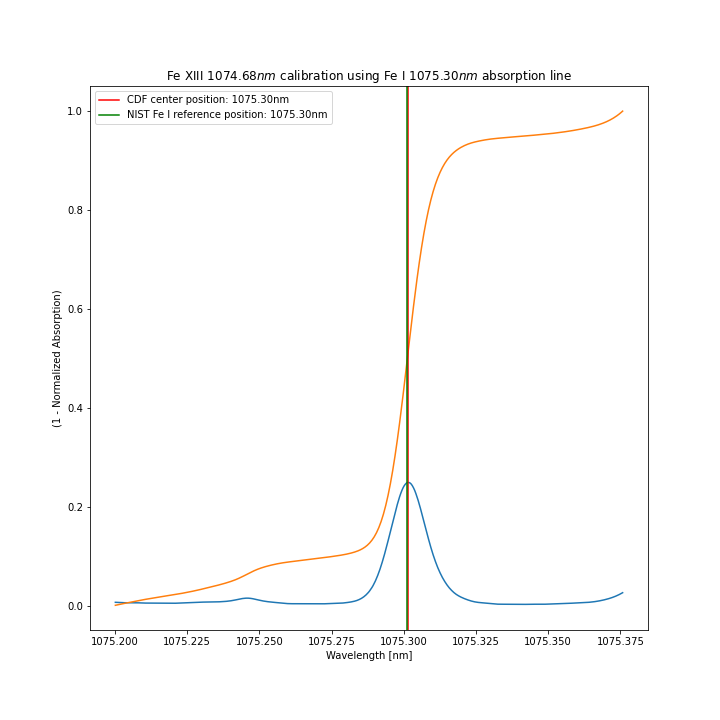}
  \includegraphics[width=0.34\linewidth]{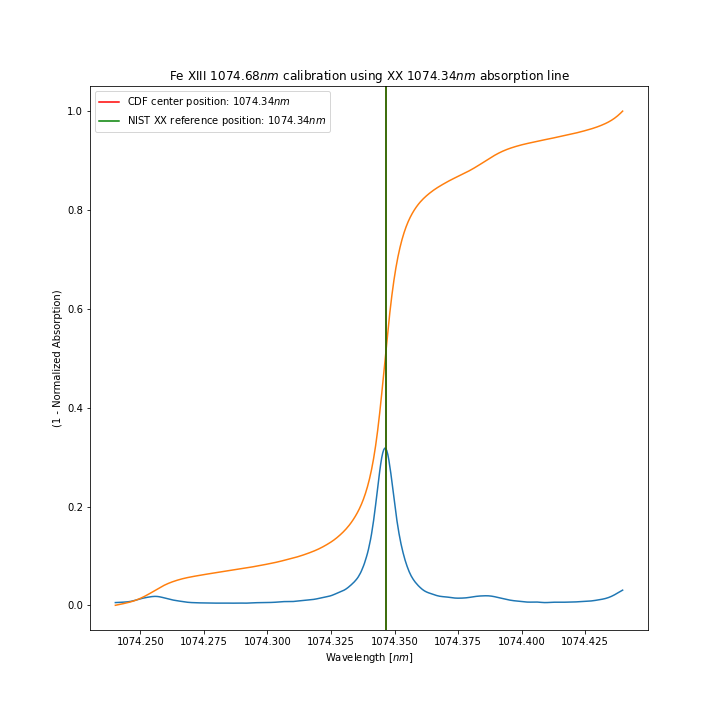}  
  \includegraphics[width=0.34\linewidth]{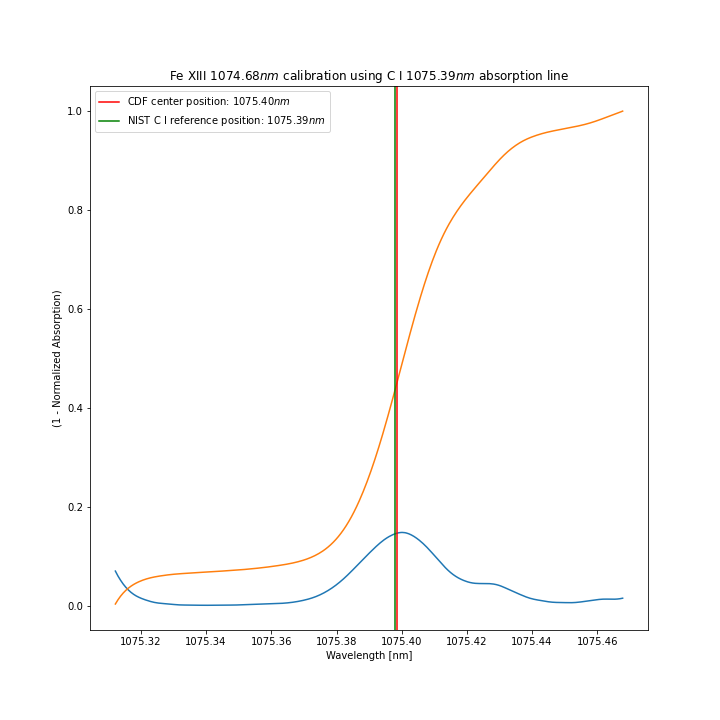}
  \includegraphics[width=0.34\linewidth]{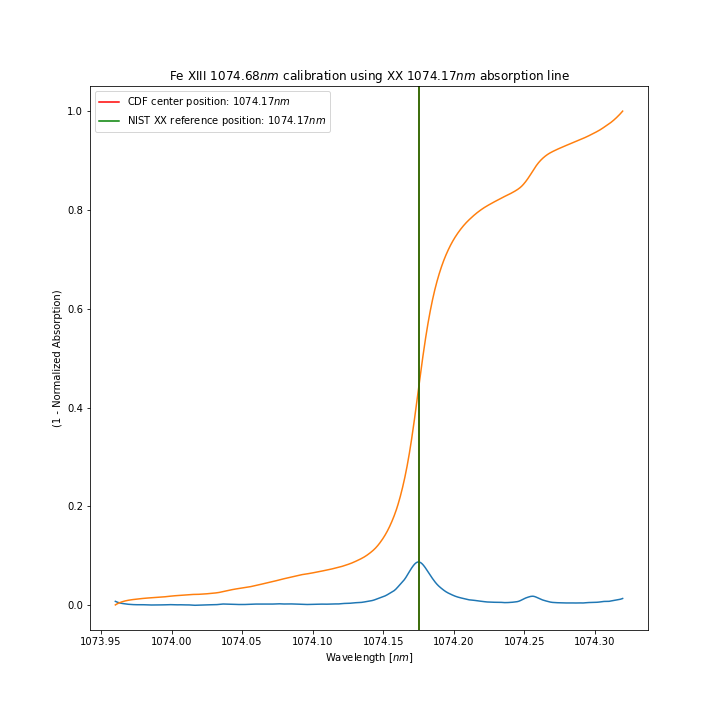}  
  \includegraphics[width=0.34\linewidth]{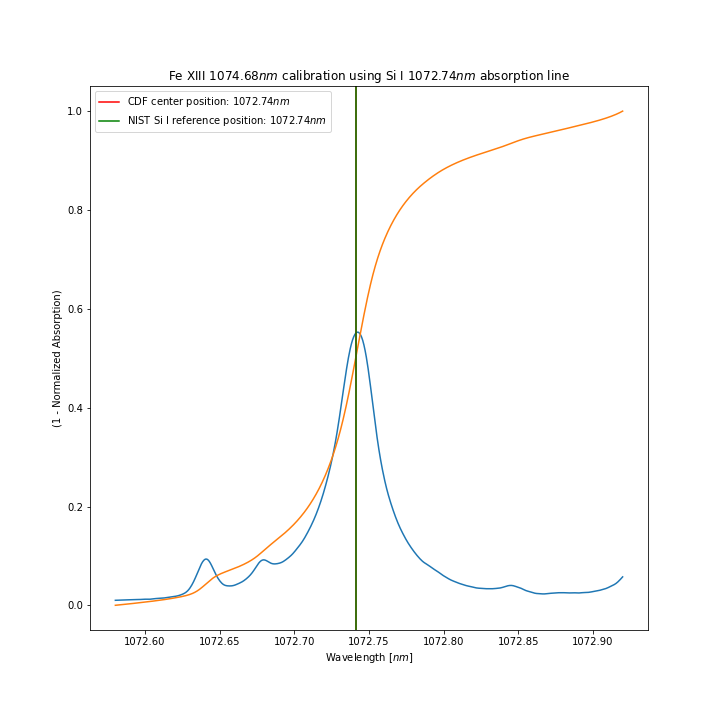} \includegraphics[width=0.34\linewidth]{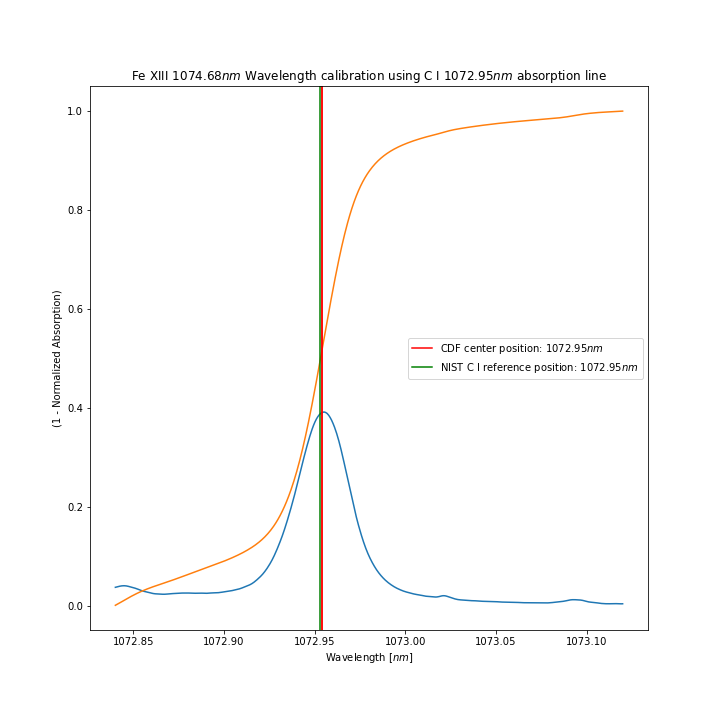}   
  \caption{\ion{Fe}{13} 1074.68 nm spectral window individual absorption line CDF fits. The A, B, C labeled lines as well as O1, O2, O3 are plotted. {The CDF counts shown in orange are normalized to a range between 0 and 1.} Note that the absorption profiles are reversed into emission to make the plot more easily readable.}
  \label{fig:appfe1}
 \end{figure}
 \FloatBarrier
 
\subsection{\ion{Fe}{13} 1079.79 nm spectral window}

\begin{figure}[h]
   \centering
   \includegraphics[angle=270,origin=c,width=0.72\linewidth]{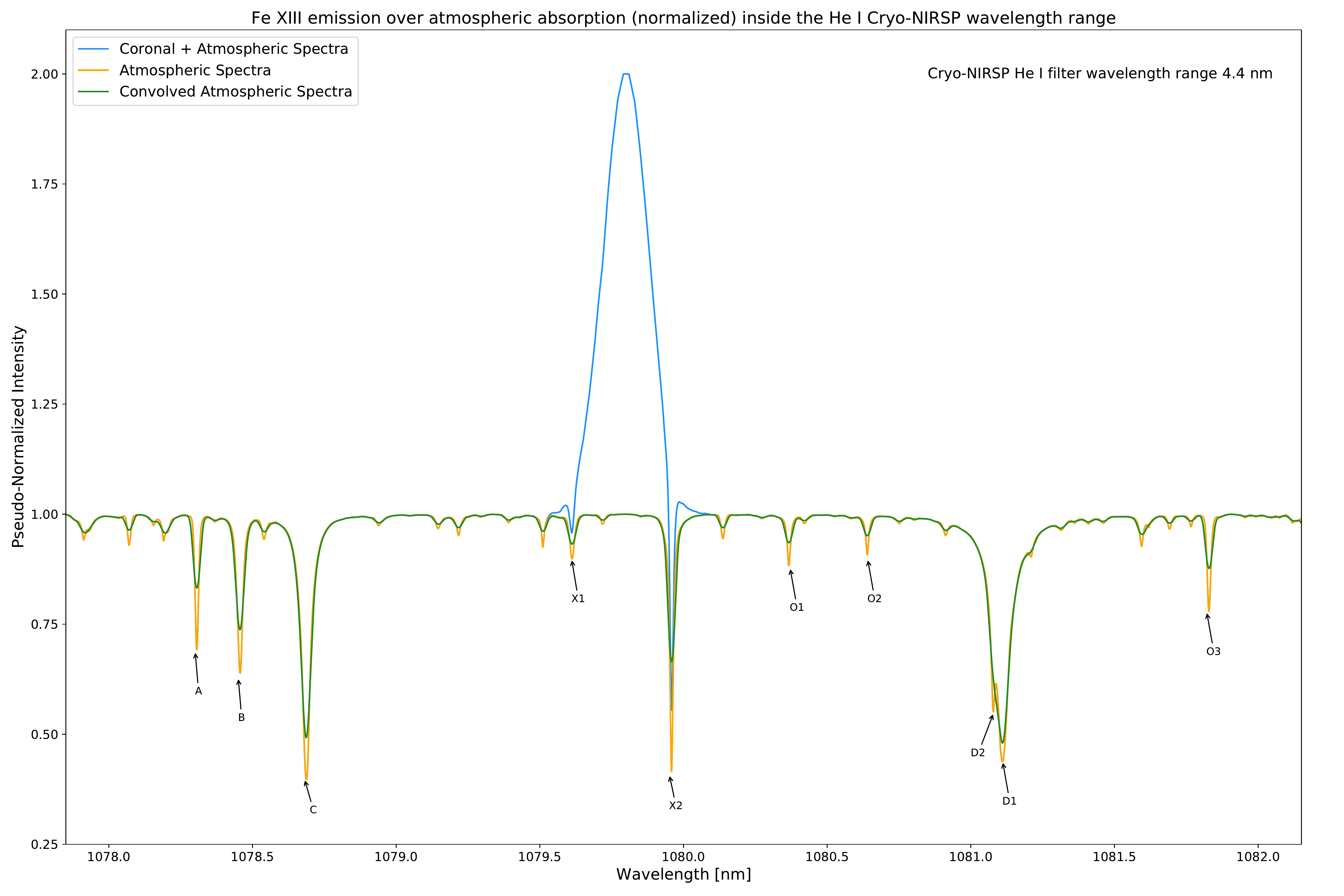}
   \caption{\ion{Fe}{13} 1079.79 nm emission over atmospheric absorption (normalized) in the He I Cryo-NIRSP wavelength range with all labeled candidate lines in Table \ref{tabtable1}. The green curve represents the atmospheric spectra that is convolved with a Gaussian function correspondent to the Cryo-NIRSP 0\farcs5 slit spectral resolution of 0.027 nm of the \ion{He}{1} filter (see \href{https://nso.edu/downloads/cryonirsp_inst_summary.pdf}{[1]}).}
   \label{fig:label7}
 \end{figure} 

\begin{figure}[!ht]
  \includegraphics[width=0.34\linewidth]{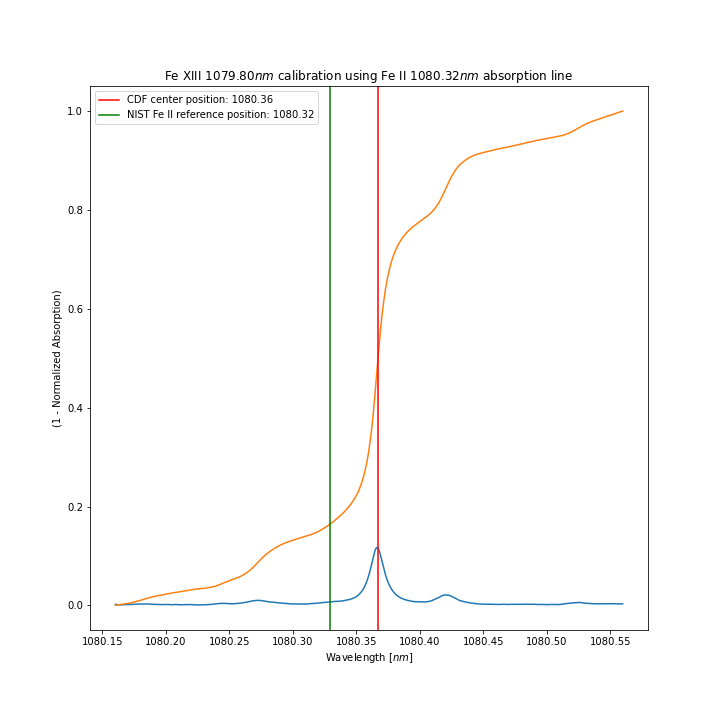}
  \includegraphics[width=0.34\linewidth]{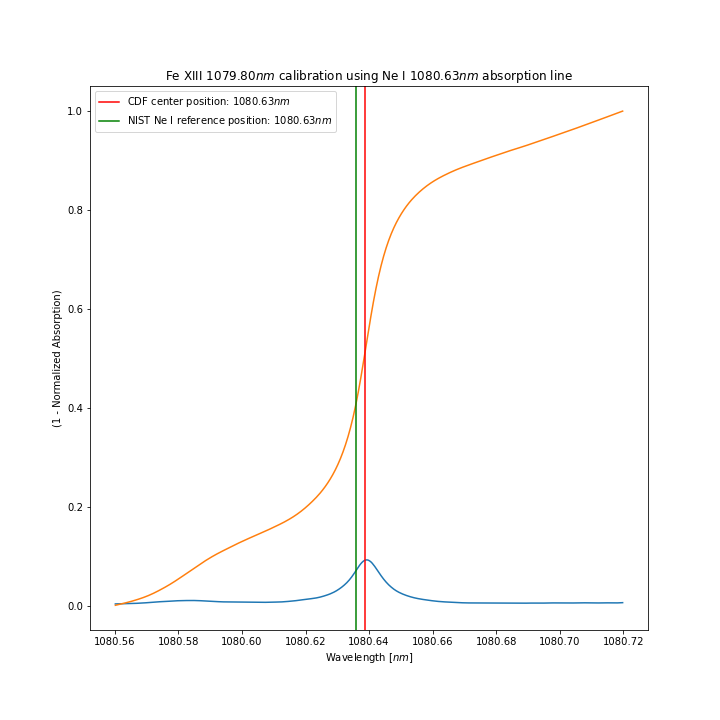}  \includegraphics[width=0.34\linewidth]{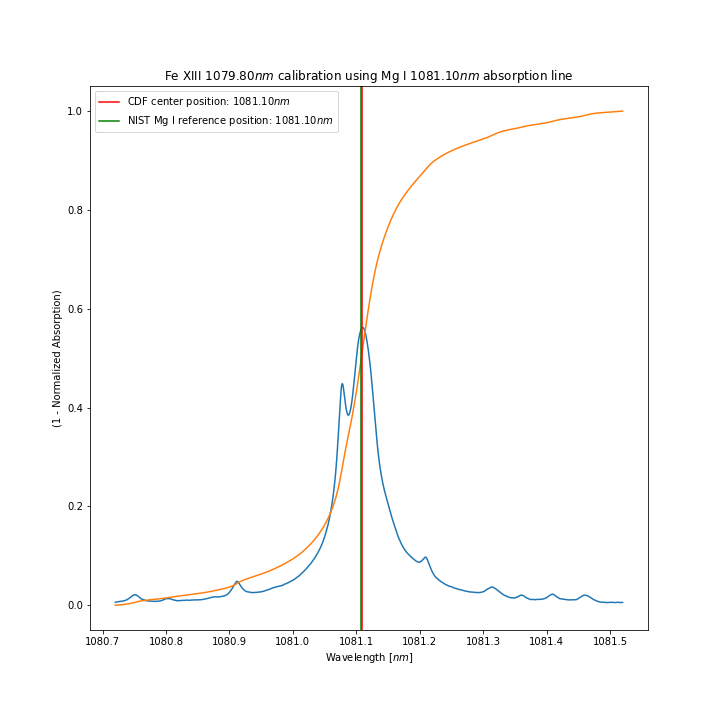}
  \includegraphics[width=0.34\linewidth]{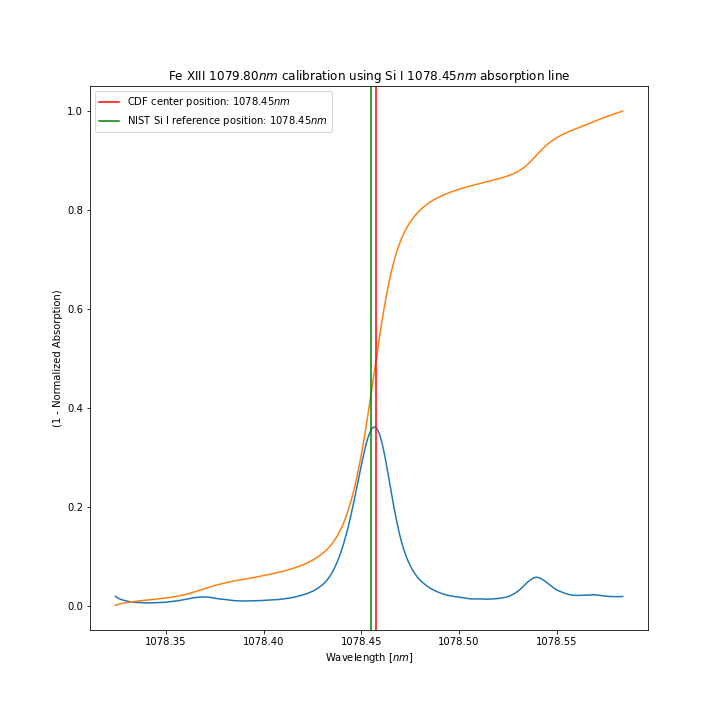}   
  \includegraphics[width=0.34\linewidth]{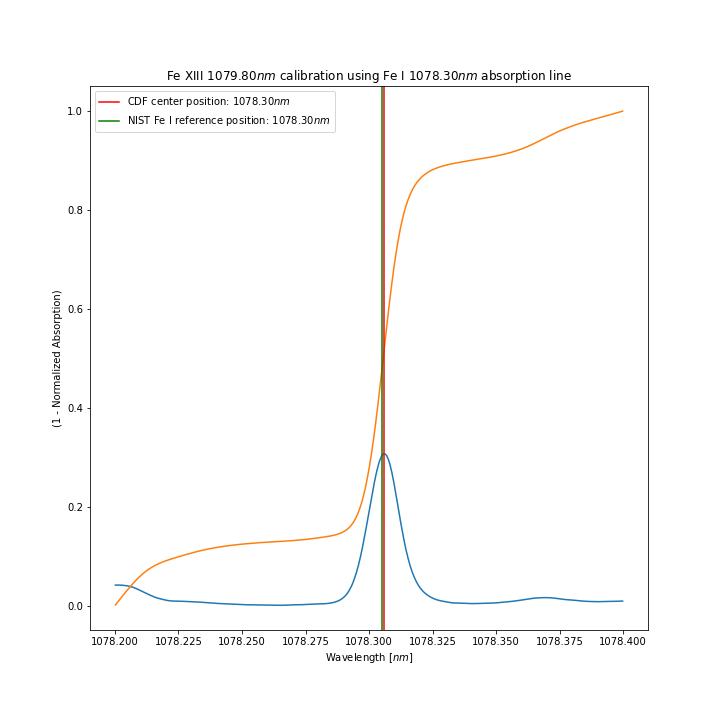}   
  \includegraphics[width=0.34\linewidth]{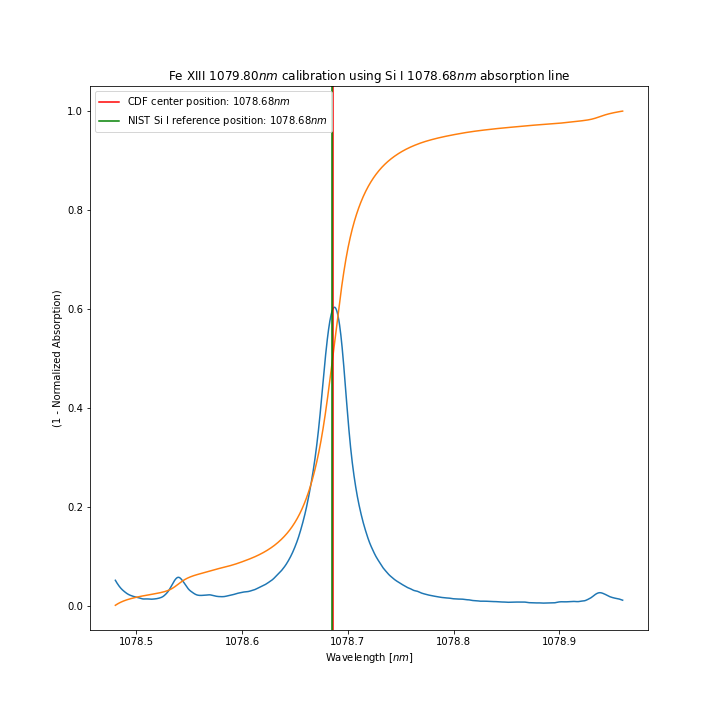}   
  \includegraphics[width=0.34\linewidth]{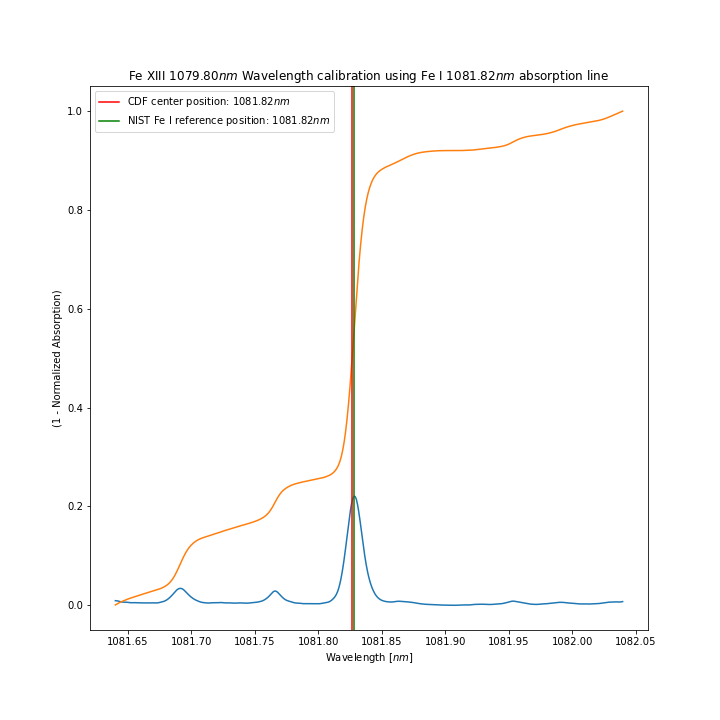}
  \caption{\ion{Fe}{13} 1079.79 nm spectral window individual absorption line CDF fits.  The A, B, C, D1 \& 2 labeled lines as well as O1, O2, O3 are plotted. {The CDF counts shown in orange are normalized to a range between 0 and 1.} Note that the absorption profiles are reversed into emission to make the plot more easily readable.}
  \label{fig:appfe2}
 \end{figure} 
 \FloatBarrier
 
~\vspace{2cm}
\subsection{\ion{Si}{10} 1430.10 nm spectral window}
 \begin{figure}[h]
   \centering
   \includegraphics[angle=270,origin=c,width=0.62\linewidth]{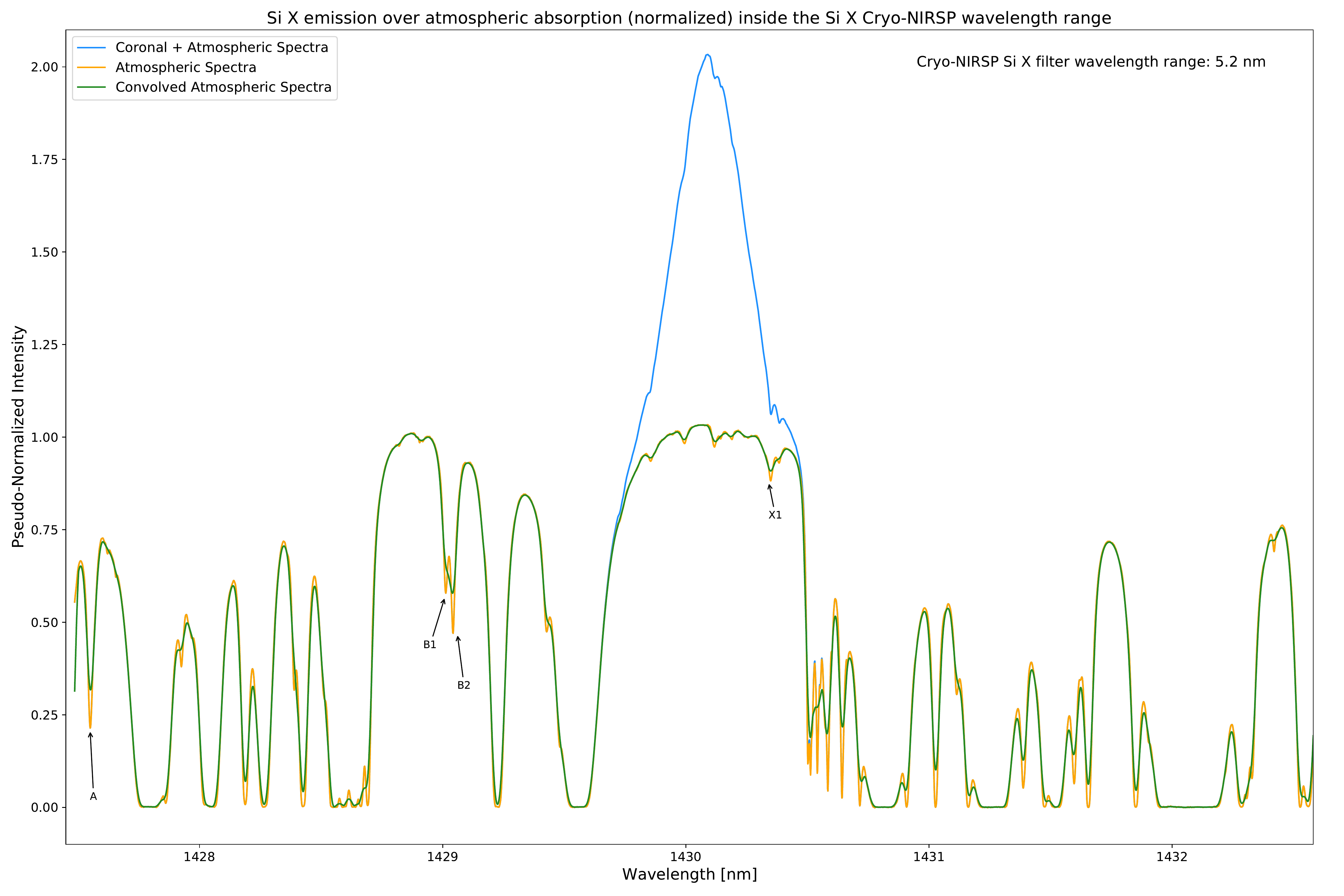}
   \caption{\ion{Si}{10} 1430.10 nm emission over atmospheric absorption (normalized) in the Si X Cryo-NIRSP wavelength range with all labeled candidate lines in Table \ref{tabtable1}. The green curve represents the atmospheric spectra that is convolved with a Gaussian function correspondent to the Cryo-NIRSP 0\farcs5 slit spectral resolution of 0.033 nm of the \ion{Si}{10} filter(see \href{https://nso.edu/downloads/cryonirsp_inst_summary.pdf}{[1]}).}
   \label{fig:label8}
 \end{figure}  
 
\begin{figure}[!hb]
  \centering
  \includegraphics[width=0.34\linewidth]{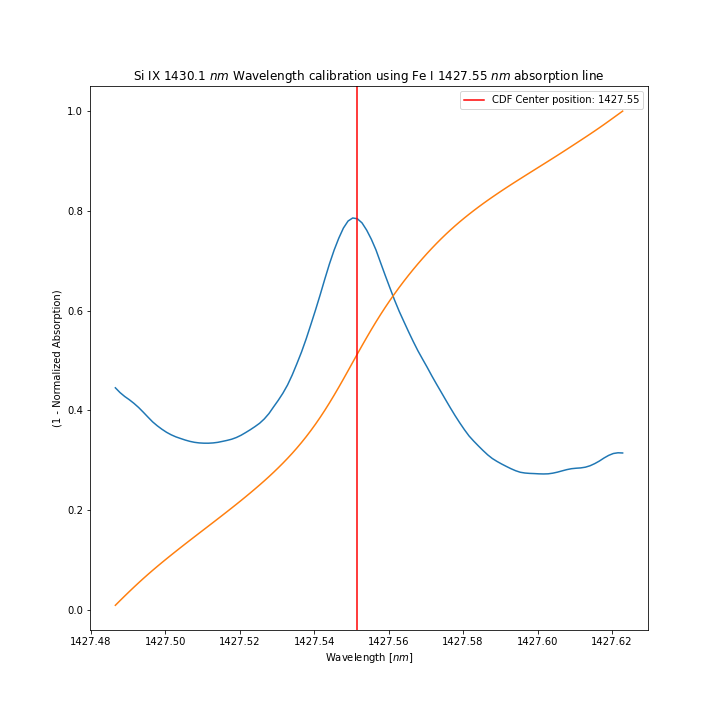}
  \includegraphics[width=0.34\linewidth]{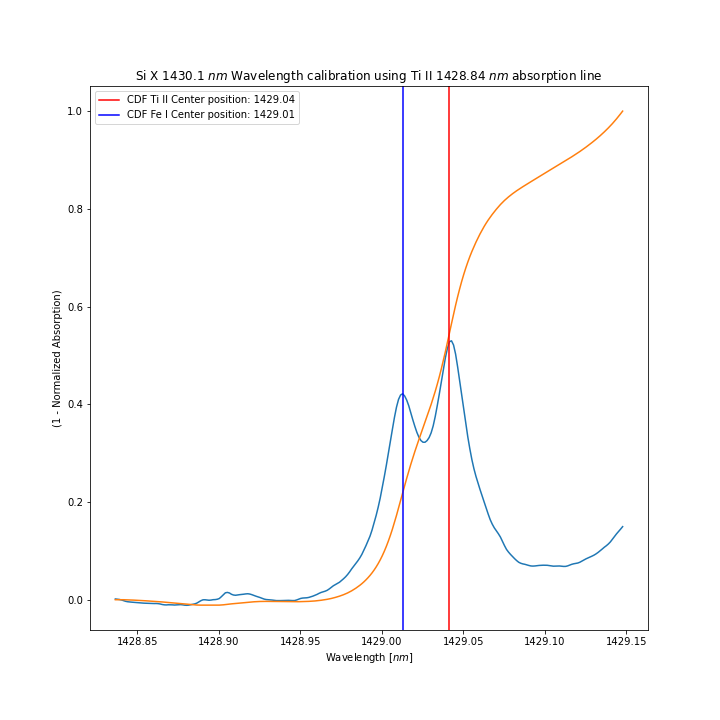}  
  \caption{\ion{Si}{10} 1430.10 nm spectral window individual absorption line CDF fits. The A, B1 \& 2 labeled lines are plotted. Only few lines proved suitable due to the convoluted spectral region. {The CDF counts shown in orange are normalized to a range between 0 and 1.} Note that the absorption profiles are reversed into emission to make the plot more easily readable. }
  \label{fig:appsi1}
 \end{figure}
  \FloatBarrier
  
 
 \subsection{\ion{Si}{9} 3934.34 nm spectral window}
 \begin{figure}[htb]
   \centering
   \includegraphics[angle=270,origin=c,width=0.75\linewidth]{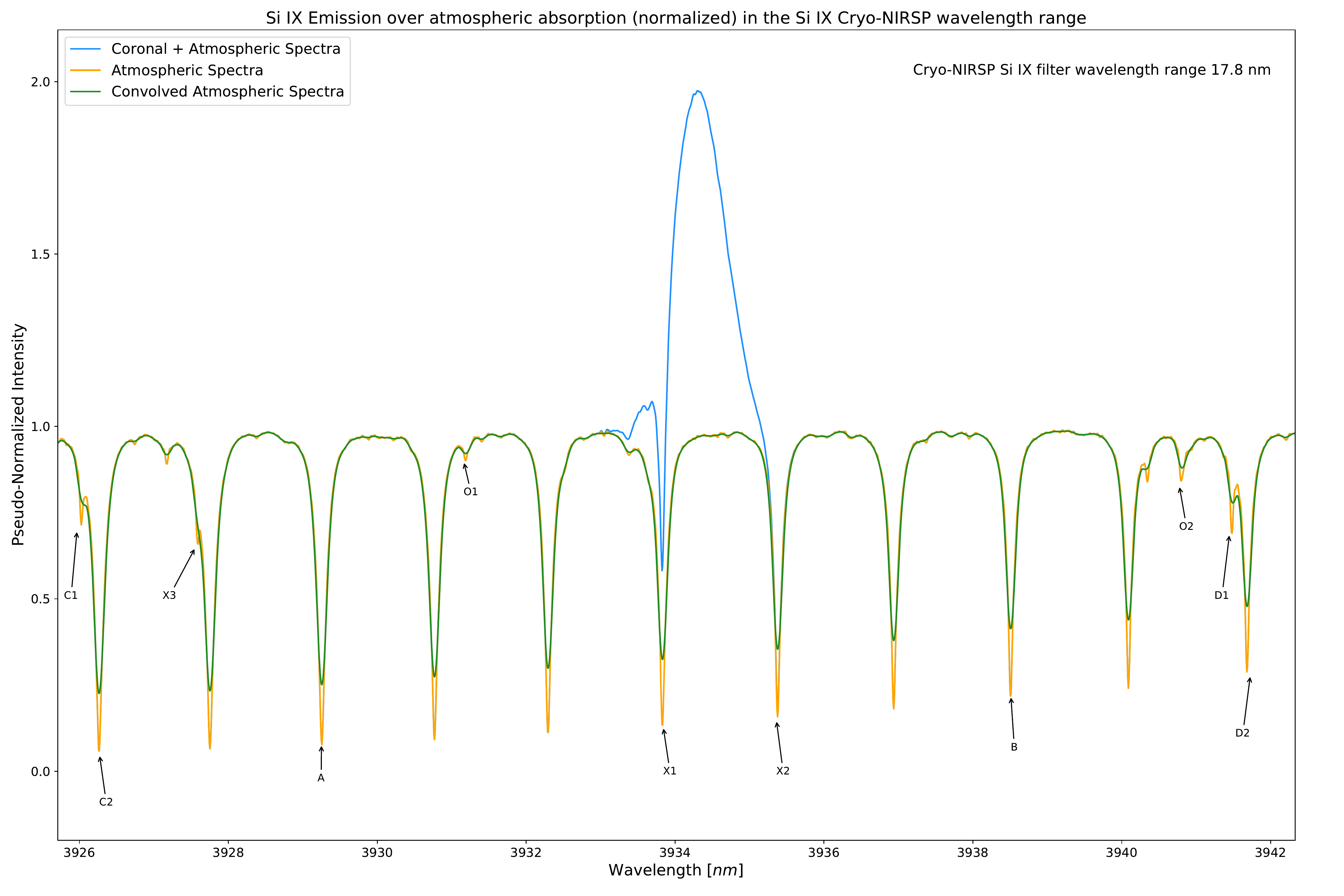}
   \caption{\ion{Si}{9} 3934.34 nm emission over atmospheric absorption (normalized) in the Si IX Cryo-NIRSP wavelength range with all labeled candidate lines in Table \ref{tabtable1}. The green curve represents the atmospheric spectra that is convolved with a Gaussian function correspondent to the Cryo-NIRSP 0\farcs5 slit spectral resolution of 0.107 nm of the \ion{Si}{9} filter (see \href{https://nso.edu/downloads/cryonirsp_inst_summary.pdf}{[1]}).}
   \label{fig:label9}
 \end{figure}

\begin{figure}[!ht]
  \includegraphics[width=0.34\linewidth]{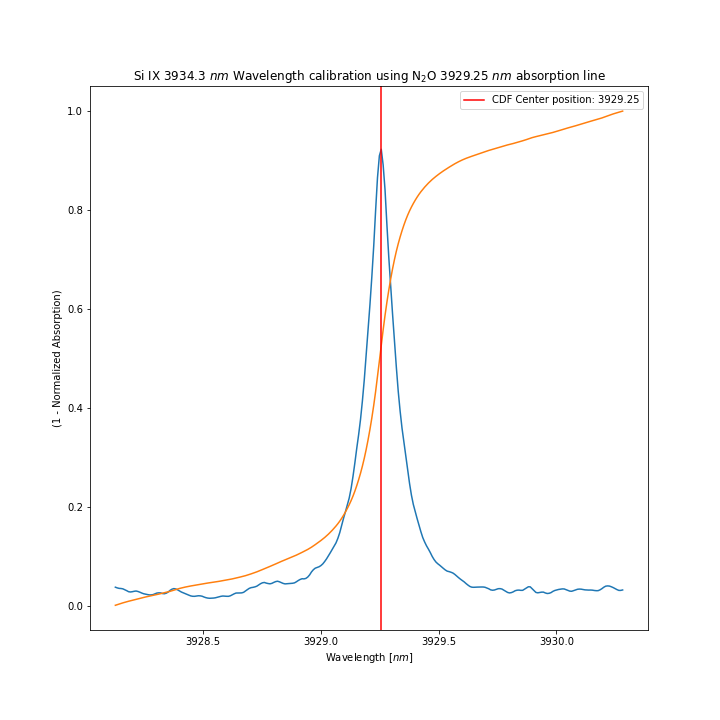}
  \includegraphics[width=0.34\linewidth]{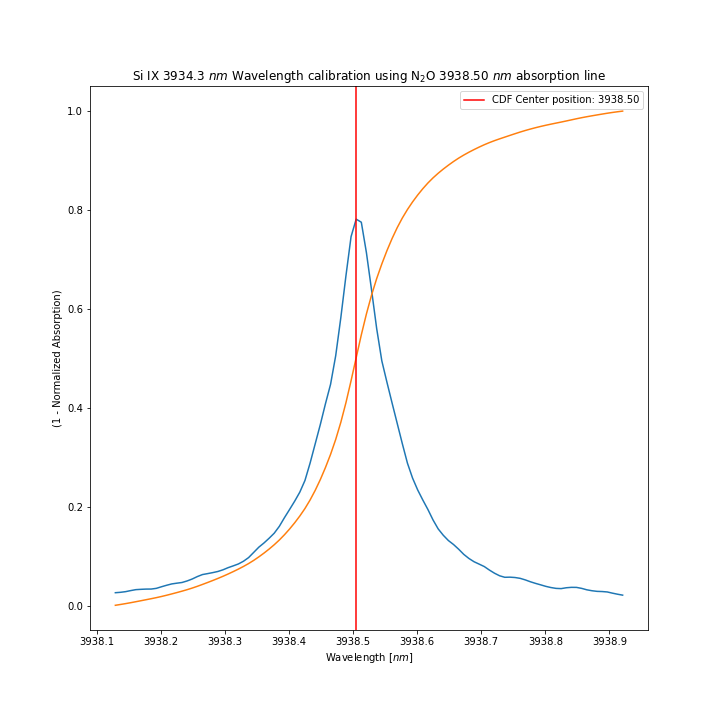} 
  \includegraphics[width=0.34\linewidth]{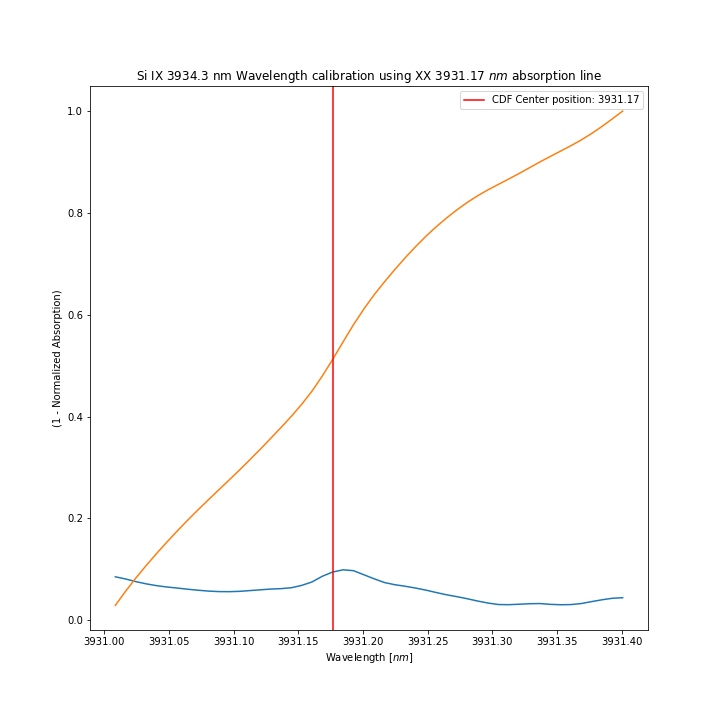}
  \includegraphics[width=0.34\linewidth]{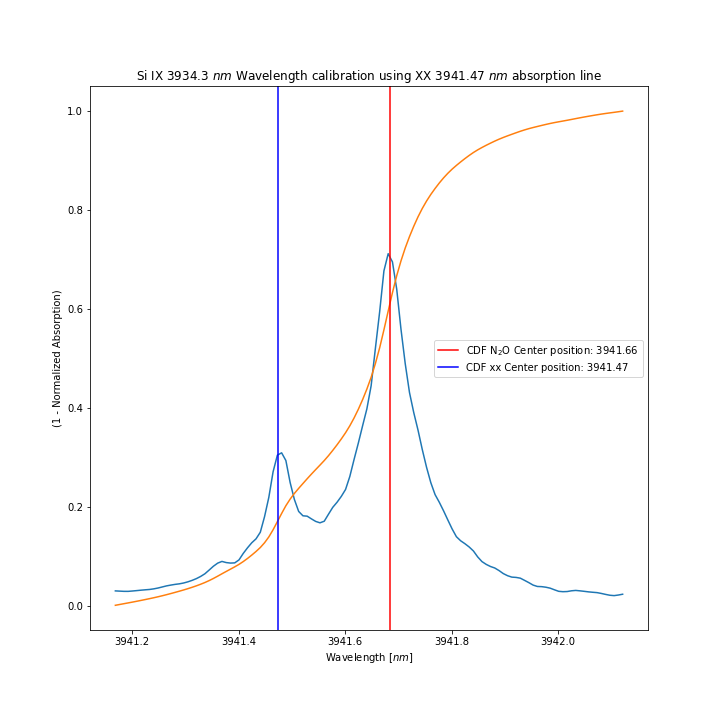}  
  \includegraphics[width=0.34\linewidth]{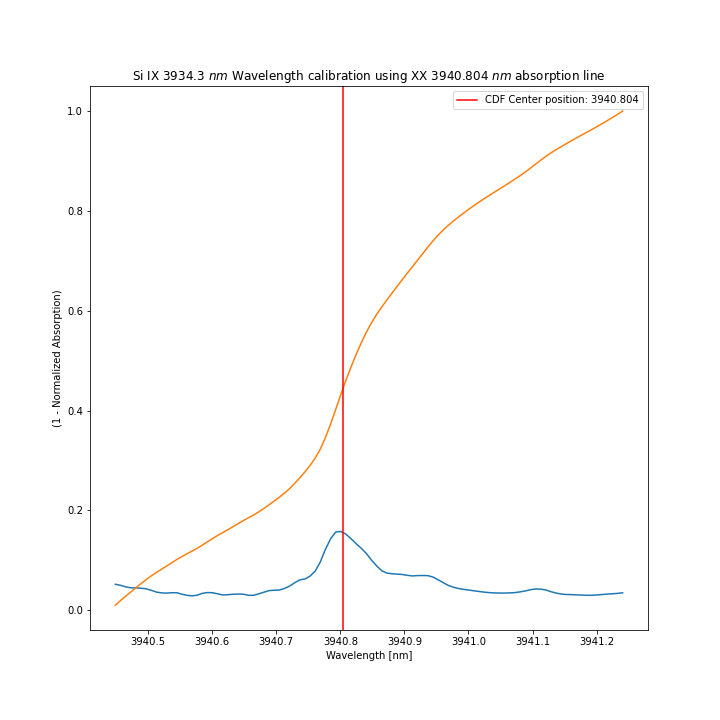} 
  \includegraphics[width=0.34\linewidth]{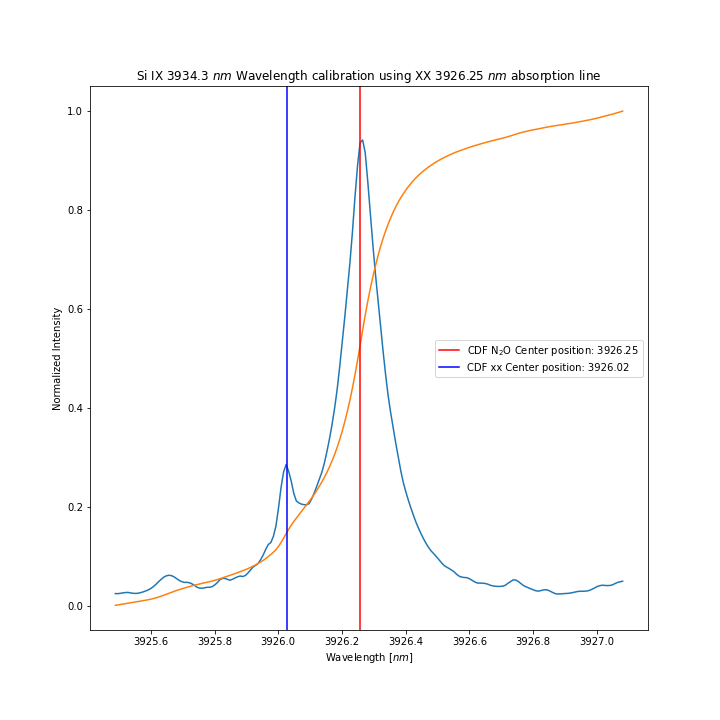} 
  \caption{\ion{Si}{9} 3934.34 nm spectral window individual absorption line CDF fits. The A, B, C1 \& 2, D1 \& 2 labeled lines as well as O1, O2 are plotted. {The CDF counts shown in orange are normalized to a range between 0 and 1.} Note that the absorption profiles are reversed into emission to make the plot more easily readable.}
  \label{fig:appsi2}
 \end{figure}
  \FloatBarrier

\end{document}